\documentclass[11pt]{article}
\usepackage[utf8]{inputenc}
\usepackage{amsmath}
\usepackage{amssymb}
\usepackage{graphicx}
\usepackage{esint}

\makeatletter

\providecommand{\tabularnewline}{\\}

\usepackage{float}\usepackage{times}\usepackage{sibgrapi}

\providecommand{\tabularnewline}{\\}
\@ifundefined{showcaptionsetup}{}{ \PassOptionsToPackage{caption=false}{subfig}}

\@ifundefined{showcaptionsetup}{}{%
 \PassOptionsToPackage{caption=false}{subfig}}
\usepackage{subfig}
\makeatother

\begin{document}
\title{A Review of Dynamic NURBS Approach}

\author{Josildo Pereira da Silva\authortag{1} and Antônio Lopes Apolinário Júnior\authortag{1} and Gilson A. Giraldi\authortag{2}}

\address{  \authortag{1} UFBA--Federal University of Bahia\\   Bahia, Brazil\\   {\tt \{josildo091@dcc.ufba.br, \{apolinario@dcc.ufba.br} \\ \\   \authortag{2}LNCC--National Laboratory for Scientific Computing\\   Av. Getulio Vargas, 333, 25651-070, Petr\'opolis, RJ, Brazil\\   {\tt \{gilson\}@lncc.br}    }

\abstract{\ Dynamic NURBS, also called  D-NURBS, is a known dynamic version of the nonuniform rational B-spline (NURBS) which integrates free-form shape representation and a physically-based model in a unified framework. More recently, computer aided design (CAD) and finite element (FEM) community realized the need to unify CAD and FEM descriptions which motivates a review of D-NURBS concepts. Therefore, in this paper we describe D-NURBS theory in the context of $1D$ shape deformations. We start with a revision of NURBS for parametric representation of curve spaces. Then, the Lagrangian mechanics is introduced in order to complete the theoretical background. Next, the D-NURBS framework for $1D$ curve spaces is presented  as well as some details about constraints and numerical implementations. In the experimental results, we focus on parameters choice and computational cost.}

\maketitle

\section{Introduction}

In the context of animation of soft objects every engine is composed
by three linked parts: the geometric model, dynamic model and rendering
module. The former can be realized in the context of parametric frameworks
like nonuniform rational B-spline (NURBS) \cite{piegl_nurbs_1997,farin_curves_1997}.
The dynamic model needs physic models that incorporate dynamic quantities
like velocity, mass and force distributions, into an evolution equation
that governs the shape deformation \cite{erleben_physics-based_2005}.
The latter includes global/local illumination techniques to generate
the scene with the desired realism \cite{pharr:04}. In this work
we focus only on the first two components.

Non-uniform Rational B-spline (NURBS) is a mathematical framework
commonly used for generating and representing curves, surfaces and
volumes \cite{piegl_nurbs_1997}. It offers an unified mathematical
basis to describe analytic and free-form shapes with great flexibility
and precision. NURBS became a standard for CAD (Computer Aided Design)
systems due to its excellent mathematical, numeric and algorithmic
properties. NURBS are built from the B-spline function basis and a
NURBS curve is a composition of NURBS functions, a set of control
points $\{\mathbf{p}_{1},\mathbf{p}_{2},\cdot\cdot\cdot,\mathbf{p}_{n}\}\subset\Re^{3}$
and a weight vector $(w_{1},w_{2},\cdot\cdot\cdot,w_{n})$. The control
points and the weights compose the degrees of freedom of the NURBS
curve.

For computer graphics applications, the dynamic model in general is
based on classical mechanics which is concerned with physical laws
to describe the behavior of a macroscopic system under the action
of forces \cite{deusen_elements_2004}. For instance, when considering
a particle in the $3D$ space under the action of gravity, we can
take its position vector along the time $t$, which in cartesian coordinates
is given by $(x(t),y(t),z(t))$, and use the Newton's laws to get
the governing equation written in terms of the cartesian coordinates
and the time $t$. In a more general situation, the instantaneous
configuration of a system may be described by the values of $n$ \emph{generalized}
coordinates $(p_{1},p_{2},\cdot\cdot\cdot,p_{n})$. So, we need a
methodology to write the evolution equation of the system in terms
of the generalized coordinates.

The Lagrangian formulation of mechanics is a framework to address
this issue \cite{goldstein_classical_1981}. It is a variational formulation
of mechanics based on the integral Hamilton's Principle which states
that the motion of the system between times $t_{1}$ and $t_{2}$
is derivable from the solution of a variational problem \cite{goldstein_classical_1981}.
The corresponding Lagrange's equations allow to write the evolution
of the system in term of the generalized coordinates. That is what
we need to link the geometric model of NURBS and the dynamic model:
we can use the control points and the weights as generalized coordinates
to describe the physical system. Therefore, we get an approach that
integrates shape representation and a dynamic model in a unified framework
called D-NURBS in the literature \cite{terzopoulos_dynamic_1994,qin_d-nurbs:_1996}.

Continuous systems, like an elastic curve, have infinite degrees of
freedom which difficult its description for both the geometric and
dynamic aspects. In mathematical terms, we are dealing with infinite
basis functions, may be uncountable. One possibility to simplify the
problem is to consider finite dimensional representation with enough
flexibility in order to represent the solution with the desired precision.
In the context of mechanical systems the Finite Element Method (FEM)
is the traditional way to perform this task. However, as pointed out
in \cite{cottrell_isogeometric_2009}, NURBS framework can be also
considered. That is way geometric modeling and FEM community realized
the need to unify CAD and FEM descriptions which motivates our review
of NURBS and D-NURBS concepts.

So, in section \ref{sec:NURBS-Rev} we start with an objective review
of B-splines functions in order to set up the background for NURBS
development. Next, in section \ref{sec:Lagragian} we describe the
Lagrangian mechanics framework in the presence of constraints and
a generalized potential for dissipation forces. Then, section \ref{sec:DNUBS-Formulation}
considers the D-NURBS model following the presentation given in \cite{terzopoulos_dynamic_1994}.
We present details of the evolution equation generation, constraints
introduction and numerical aspects. For simplicity, we focus on curve
spaces but the theory can be straightforward generalize for surfaces
and volumes. In the experimental results (section \ref{sec:Exper}),
we consider a set up for a linear mass distribution with fixed end
points. We discuss the influence of parameters choice, effects of
NURBS weights and computational cost. The conclusions and further
works are presented in section \ref{sec:Concl}, The appendices \ref{appendix:A}
and \ref{appendix:B} give some details about specific terms of the
D-NURBS governing equation.

\section{NURBS: Nonuniform Rational B-spline \label{sec:NURBS-Rev}}

The spline framework is the starting point for NURBS development.
A polynomial spline of order $k$ (degree $k-1$) is a piecewise polynomial
function of order $k$ with continuity of derivative of order $k-2$
at the common joints between segments, which are called \emph{patches}
\cite{rogers_mathematical_1976,persiano_bases_1996}.

Therefore, the spline space is a functional space composed by piecewise
polynomial functions with the property already stated. A fundamental
element in the spline theory is the \emph{knot vector} which defines
the end points of the patches of the spline function. Given the order
$k$ and a knot vector $\mathbf{v}=(u_{0},u_{1},\cdot\cdot\cdot,u_{n})$,
we can denote the space of polynomial splines of order $k$ with domain
in the range $[u_{0},u_{n}]$ as $S^{k}(u_{0},u_{1},\cdot\cdot\cdot,u_{n})$.
The Figure \ref{fig:spline-examples} shows some elements of this
set when $k=1,2,3$.

\begin{center}
\begin{figure}[htb]
\begin{centering}
\includegraphics[scale=0.8]{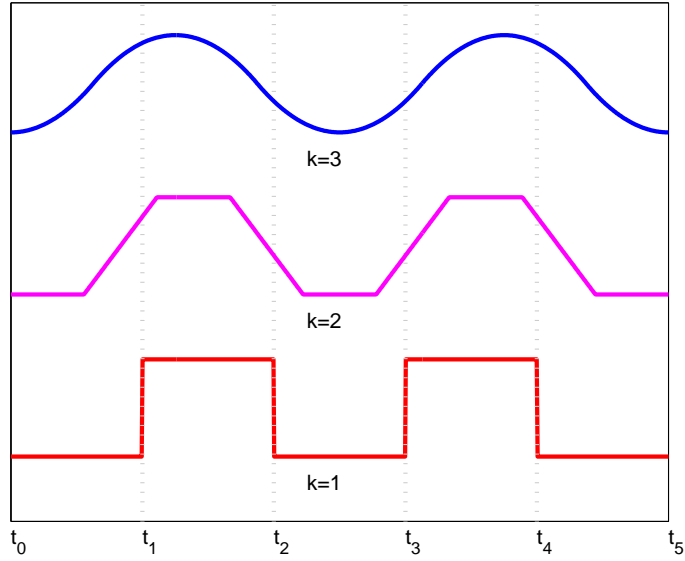} 
\par\end{centering}

\caption{Polynomial spline examples with knot vector $\mathbf{v}=(t_{0},t_{1},t_{2},t_{3},t_{4},t_{5})$.}

\label{fig:spline-examples} 
\end{figure}

\par\end{center}

We can show that the $S^{k}(u_{0},u_{1},\cdot\cdot\cdot,u_{n})$ is
a vector space of dimension $n-k+1$ \cite{persiano_bases_1996}.
The main point in the spline theory is to construct a basis for this
space. From the functional analysis viewpoint the space properties
are invariant respect to the basis choice. However, for computer graphics
aspects it is important that every tool and algorithm generated has
an intuitive geometric and visual interpretation with local control
of the target objects. The B-spline basis attend these requirements.

Following traditional texts in this area \cite{farin_curves_1997,rogers_mathematical_1976}
we perform a recursive definition of the B-spline basis. So, let us
consider the $S^{1}(u_{0},u_{1},\cdot\cdot\cdot,u_{n})$; that means,
the space of piecewise polynomial functions of order $k=1$ (degree
$0$) which are just piecewise constant functions, like the one presented
on Figure \ref{fig:spline-examples} for $k=1$.

A basis for this space is in fact the first B-spline basis in our
recursive scheme, which is defined as follows:

\begin{equation}
B_{i,1}(u)=\begin{cases}
1, & if\: u_{i}\leq u<u_{i+1}\\
0, & otherwise
\end{cases}\label{bspline-order0}
\end{equation}
for $i=0,1,\cdot\cdot\cdot,n-1$.

Now, let us consider the space $S^{2}(u_{0},u_{1},\cdot\cdot\cdot,u_{n});$
that is, the space of piecewise polynomial functions of order $k=2$
(degree $k-1=1$) which are just piecewise linear functions with continuity
of derivative of order $k-2=0$ (see Figure \ref{fig:spline-examples}).
We are supposing that $u_{0}<u_{1}<\cdot\cdot\cdot<u_{n}$. We already
know that this is a vector space with dimension $n-k+1=n-1.$ Besides,
when integrating polynomial functions of order $k$ we get again polynomial
functions but with order $k+1.$ Also, we want that the support of
the functions $B_{i,2}$ would be as small as possible (for local
geometric control) and that they have continuity of derivative of
order $k-2=0$ at the common joints between patches. The following
functions fulfill these requirements:

\begin{equation}
B_{i,2}\left(u\right)=\int_{-\infty}^{u}\left(\frac{B_{i,1}\left(s\right)}{u_{i+1}-u_{i}}-\frac{B_{i+1,1}\left(s\right)}{u_{i+2}-u_{i+1}}\right)ds,\quad i=0,1,\cdot\cdot\cdot,n-2.
\end{equation}

By repeating the above arguments, we can show that the following recursive
scheme will generate a basis $B^{k}=\left\{ B_{i,k},\quad i=0,1,\cdot\cdot\cdot,n-k-1\right\} $
for splines $f:\left[u_{0},u_{n}\right]\rightarrow\Re$ such that
$f\left(u_{0}\right)=f\left(u_{n}\right)=0$:

\begin{equation}
B_{i,k}\left(u\right)=\frac{\left(u-u_{i}\right)B_{i,k-1}\left(u\right)}{u_{i+k-1}-u_{i}}+\frac{\left(u_{i+k}-u\right)B_{i+1,k-1}\left(u\right)}{u_{i+k}-u_{i+1}},\quad i=0,1,\cdot\cdot\cdot,n-k\label{bspline01}
\end{equation}
where $k=2,3,\cdot\cdot\cdot$ and the $B_{i,1}\left(u\right)$ is
given by expression (\ref{bspline-order0}). The Figure \ref{fig:Bsplines-degree1}
pictures the obtained basis for $k=2.$

\begin{center}
\begin{figure}[htb]
\begin{centering}
\includegraphics[scale=0.8]{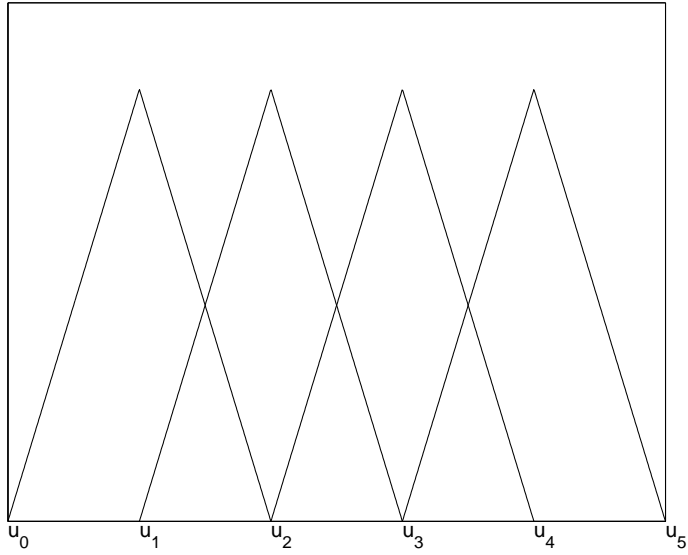} 
\par\end{centering}

\caption{B-spline of order $k=2$ with knot vector $\mathbf{v}=(u_{0},u_{1},u_{2},u_{3},u_{4},u_{5})$.}

\label{fig:Bsplines-degree1} 
\end{figure}

\par\end{center}

So, in the above development, the span of $B^{k}$ is in fact a subspace
of $S^{k}(u_{0},u_{1},\cdot\cdot\cdot,u_{n})$ once $B^{k}$ can only
generate functions with support in the interval $(u_{0},u_{n})$ as
we already observed above. However, we can cover all the spline space
by considering more general knot vectors. In fact, the knot vector
has a significant influence in the spline basis generated. In general,
it is used three types of knot vectors: uniform, open uniform (or
just open) and nonuniform.

Uniform knot vectors satisfies $u_{i+1}-u_{i}=\Delta u=const.,$ for
$i=1,2,...,n.$ Uniform knot vectors yield periodic uniform basis
functions, like the one presented in Figure \ref{fig:Bsplines-uniform-degree1};
that means:

\[
B_{i,k}\left(u\right)=B_{i-1,k}\left(u-\Delta u\right)=B_{i+1,k}\left(u+\Delta u\right).
\]

\begin{center}
\begin{figure}[htb]
\begin{centering}
\includegraphics[scale=0.8]{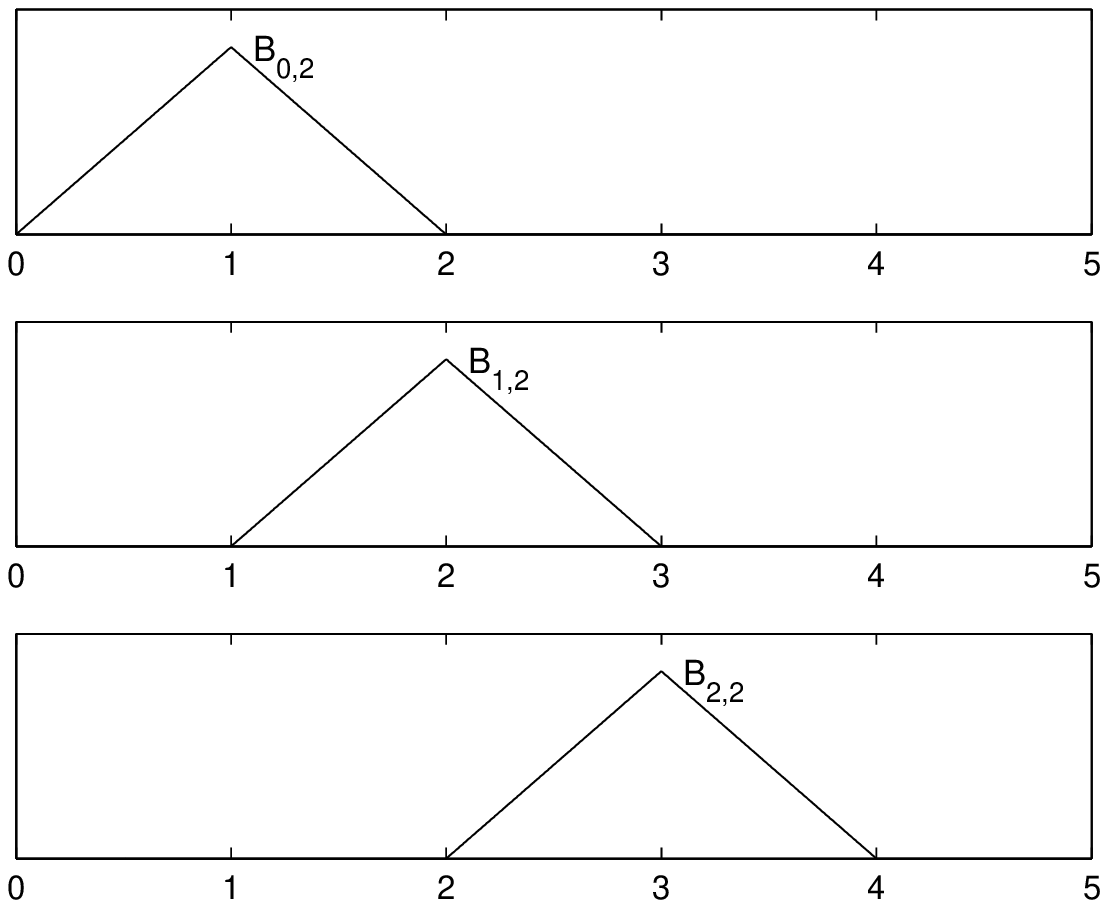} 
\par\end{centering}

\caption{B-splines examples of order $k=2$ with uniform knot vector $\mathbf{v}=(0,1,2,3,4,5)$.}

\label{fig:Bsplines-uniform-degree1} 
\end{figure}

\par\end{center}

An open uniform knot vector has also the property $u_{1+1}-u_{i}=\Delta u$
for internal knots but it has multiplicity of knot values at the ends
equal to the order $k$ of the B-spline functions. For instance:

\[
\mathbf{v}=(0,0,1,2,3,4,5,5),\text{ }if\quad k=2,
\]

\[
\mathbf{v}=(0,0,0,0.1,0.2,0.3,0.4,0.5,0.5,0.5),\text{ }if\quad k=3.
\]

These kind of knot vectors may yield more general B-spline basis $B^{k}$
that can generate functions that are not null at the ends of the knot
vector, as we can visualize in Figure \ref{fig:bsplines-order2-open}.

\begin{center}
\begin{figure}[htb]
\begin{centering}
\includegraphics[scale=0.8]{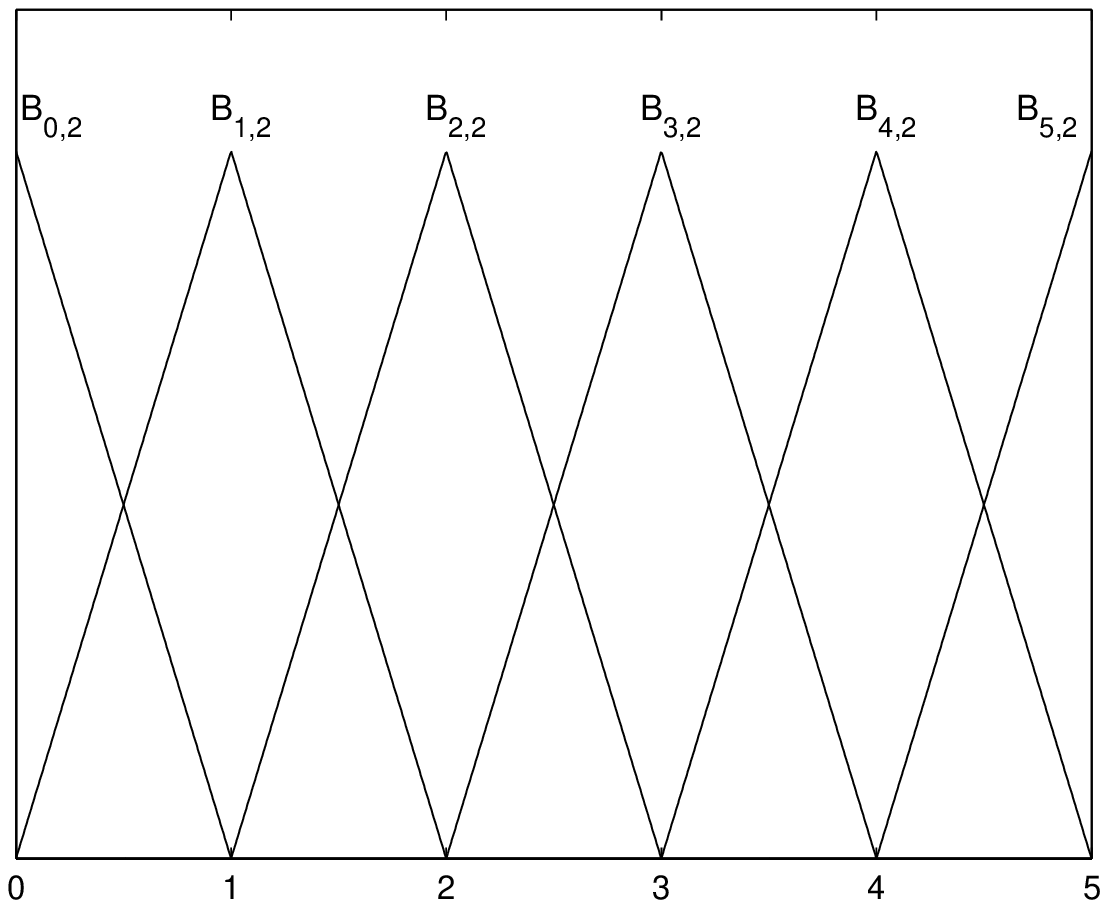} 
\par\end{centering}

\caption{B-splines functions of order $k=2$ with open uniform knot vector
$\mathbf{v}=(0,0,1,2,3,4,5,5)$. }

\label{fig:bsplines-order2-open} 
\end{figure}

\par\end{center}

Finally, nonuniform knot vectors may have either unequally spaced
($u_{1+1}-u_{i}=\Delta u_{i}$) and/or multiple knot values at the
ends or even for the internal knots.

The B-splines generated by open (uniform or nonuniform) knot vectors
have important properties \cite{piegl_nurbs_1997}.
\begin{enumerate}
\item $B_{i,k}\left(u\right)\geq0$ $\forall u$.
\item $B_{i,k}\left(u\right)=0$ if $u$ is outside the interval $\in\lbrack u_{i},u_{i+k+1})$.
\item Partition of unity: $\sum_{i=0}^{n-k}B_{i,k}\left(u\right)=1$. 
\end{enumerate}
Once defined the basis for the spline space, we can consider curve
spaces in $\Re^{3}$ generated through B-splines. So, let us take
a set of $\mathbf{p}_{i},$ $i=0,1,2,\cdot\cdot\cdot,n-k$ points
in $\Re^{3}$ and the vector-valued function given by:

\begin{equation}
\mathbf{c}\left(u\right)=\sum_{i=0}^{n-k}\mathbf{p}_{i}B_{i,k}\left(u\right).\label{spline-curve00}
\end{equation}

This function defines a curve of class $C^{k-2}$ in \ $\Re^{3},$
which is called a spline curve. The points $\mathbf{p}_{i}$ are called
control points and the corresponding polygon is the defining polygon.
Important properties about these curves are:
\begin{enumerate}
\item End points interpolation: in the case of open knot vector we have
$\mathbf{c}\left(u_{0}\right)=\mathbf{p}_{0}$ and $\mathbf{c}\left(u_{n}\right)=\mathbf{p}_{n}$.
\item Affine Invariance: If $\psi\left(r\right)=Ar+v$ is an affine transformation
then $\psi\left(c\left(u\right)\right)=\sum_{i=0}^{n-k}B_{i,k}\left(u\right)\psi\left(\mathbf{p}_{i}\right)$.
\item Strong convex hull property: the curve belongs to the convex hull
of its control polygon. 
\end{enumerate}
A rational B-spline curve is the projection of a polynomial B-spline
curve defined in the four-dimensional homogeneous coordinate space
back into the three-dimensional physical space \cite{rogers_mathematical_1976}.
Therefore, if we represent the control points in the four-dimensional
homogeneous coordinate space we obtain:

\[
\widetilde{\mathbf{p}}_{i}=\left(\begin{array}{c}
w_{i}\mathbf{p}_{i}\\
w_{i}
\end{array}\right),\quad i=0,1,2,\cdot\cdot\cdot,n-k,
\]
and applying expression (\ref{spline-curve00}) we get a spline curve
in the four-dimensional homogeneous space:

\begin{equation}
\widetilde{\mathbf{c}}\left(u\right)=\sum_{i=0}^{n-k}\left(\begin{array}{c}
w_{i}\mathbf{p}_{i}\\
w_{i}
\end{array}\right)B_{i,k}\left(u\right).\label{nurbs00}
\end{equation}

By projection in the three-dimensional space we obtain the rational
curve:

\begin{equation}
\mathbf{c}\left(u\right)=\frac{\sum_{i=0}^{n-k}\mathbf{p}_{i}w_{i}B_{i,k}\left(u\right)}{\sum_{j=0}^{n-k}w_{j}B_{j,k}\left(u\right)}=\sum_{i=0}^{n-k}\mathbf{p}_{i}N_{i,k}\left(u\right),\label{nurbs01}
\end{equation}
where $N{}_{i,k}$ are the rational B-spline functions given by:

\begin{equation}
N{}_{i,k}\left(u\right)=\frac{w_{i}B_{i,k}\left(u\right)}{\sum_{j=0}^{n-k}w_{j}B_{j,k}\left(u\right)}.\label{rational-spline00}
\end{equation}

If the B-splines in expression \ref{bspline01} are generated by nonuniform
knot vectors then the functions $R_{i,k}$ in expression (\ref{rational-spline00})
are named nonuniform rational B-splines (NURBS) and the curve defined
by expression (\ref{nurbs01}) is a NURBS curve.

B-splines can be enriched without modifying the underlying geometry
and parameterization through the mechanisms that are called refinements.
The most common mechanisms are knot insertion and degree elevation
\cite{farin_curves_1997,piegl_nurbs_1997}.

\section{Lagrangian Mechanics \label{sec:Lagragian}}

Let us consider a physical system whose instantaneous configuration
may be described by the values of $n$ generalized coordinates $(p_{1},p_{2},\cdot\cdot\cdot,p_{n})$
which can be considered as a point in a $n-dimensional$ Cartesian
hyperspace known as \emph{configuration space}. As time goes on from
a time $t_{1}$ to a time $t_{2}$, the system changes its configuration
due to internal and external forces. Therefore, the evolution of the
system can be seem as a continuous path, or curve, $\mathbf{p}(t)$
in the, configuration space, parameterized through the time $t$.

The Hamilton's Principle gives a methodology to write the evolution
equation of the system in terms of the generalized coordinates and
time $t$. It states that if for a mechanical systems with kinetic
energy $T=T\left(\overset{\cdot}{\mathbf{p}}\right),$ where $\overset{\cdot}{\mathbf{p}}=d\mathbf{p/}dt,$
all force fields are derivable from a scalar potential $V=V\left(\mathbf{p},t\right)$
then the motion of the system from time $t_{1}$ to time $t_{2}$
is such that the line integral:

\begin{equation}
I=\int_{t_{1}}^{t_{2}}L\left(\mathbf{p,}\overset{\cdot}{\mathbf{p}},t\right)dt,\label{Lagran00}
\end{equation}
where $L\left(\mathbf{p,}\overset{\cdot}{\mathbf{p,}t}\right)=T\left(\overset{\cdot}{\mathbf{p}}\right)-V\left(\mathbf{p},t\right),$
has a stationary value for the correct path of the motion \cite{goldstein_classical_1981}.

The function $L$ is named the Lagrangian of the system and we can
apply traditional techniques of the variational calculus to show that
the correct path must satisfies:

\begin{equation}
\frac{d}{dt}\left(\frac{\partial L}{\partial\overset{\cdot}{p_{i}}}\right)-\frac{\partial L}{\partial p_{i}}=0,\quad i=1,2,\cdot\cdot\cdot,n,\label{Lagran01}
\end{equation}
or, in a compact form:

\begin{equation}
\frac{d}{dt}\left(\frac{\partial L}{\partial\overset{\cdot}{\mathbf{p}}}\right)-\frac{\partial L}{\partial\mathbf{p}}=0,\quad i=1,2,\cdot\cdot\cdot,n,\label{Lagran00-compact}
\end{equation}
which are the Lagrange equations of motion \cite{goldstein_classical_1981}.

We can introduce dissipation forces in the Hamilton' principle by
adding a velocity-dependent term in the scalar potential of the system.
So, let us consider the general form for the Lagrangian: 
\begin{equation}
L\left(\mathbf{p,}\overset{\cdot}{\mathbf{p}},t\right)=T\left(\mathbf{p,}\overset{\cdot}{\mathbf{p}}\right)-\left(U\left(\mathbf{p}\right)+F\left(\mathbf{p,}\overset{\cdot}{\mathbf{p}}\right)\right),\label{Lagrangeana00}
\end{equation}
where, like before, $T$ is the kinetic energy but now possibly dependent
from both $\mathbf{p}$ and $\overset{\cdot}{\mathbf{p}},$ $U$ is
the potential related to the conservative forces and $F$ is a velocity-dependent
potential to account for dissipative effects.

So, substituting this expression in the Euler-Lagrange equations (\ref{Lagran00-compact})
renders:

\begin{equation}
\frac{d}{dt}\left(\frac{\partial T}{\partial\overset{\cdot}{\mathbf{p}}}-\frac{\partial F}{\partial\overset{\cdot}{\mathbf{p}}}\right)-\left(\frac{\partial T}{\partial\mathbf{p}}-\frac{\partial U}{\partial\mathbf{p}}-\frac{\partial F}{\partial\mathbf{p}}\right)=0\label{Geral-Euler-Lagrange-01}
\end{equation}

In general, mechanical systems undergoes effects of internal and external
forces. Therefore, it is useful to decompose the potential $U\left(\mathbf{p}\right)$
into two terms named $E_{int}$ e $E_{ext}$, which will account for
the internal and external forces, respectively: 
\begin{equation}
U\left(\mathbf{p}\right)=E_{int}\left(\mathbf{p}\right)+E_{ext}\left(\mathbf{p}\right).\label{conservative-potential}
\end{equation}
By substituting expression (\ref{conservative-potential}) into the
equations (\ref{Geral-Euler-Lagrange-01}), we get: 
\begin{equation}
\frac{d}{dt}\left(\frac{\partial T}{\partial\overset{\cdot}{\mathbf{p}}}-\frac{\partial F}{\partial\overset{\cdot}{\mathbf{p}}}\right)+\left(\frac{\partial E_{int}}{\partial\mathbf{p}}\right)=-\frac{\partial E_{ext}}{\partial\mathbf{p}}+\left(\frac{\partial T}{\partial\mathbf{p}}-\frac{\partial F}{\partial\mathbf{p}}\right),\label{Geral-Euler-Lagrange-04}
\end{equation}
which gives the general form of Euler-Lagrange equations.

\subsection{Lagrange Equations with Constraints \label{subsec:constraints}}

Now, let us extend the Hamilton' principle in order to cover constraints.
We focus on holonomic constraints; or holonomic system, for which
the constraints may be expressed by:

\begin{eqnarray}
f_{1}(p_{1},p_{2},\cdot\cdot\cdot,p_{n},t) & = & 0,\notag\\
f_{2}(p_{1},p_{2},\cdot\cdot\cdot,p_{n},t) & = & 0,\notag\\
 &  & \cdot\cdot\cdot\label{constraint00}\\
f_{m}(p_{1},p_{2},\cdot\cdot\cdot,p_{n},t) & = & 0,\notag
\end{eqnarray}
where $f_{l},$ $l=1,2,\cdot\cdot\cdot,m,$\ is a general expression
connecting the generalized coordinates. In this case, we can take
the differential $df_{l}$:

\begin{equation}
df_{l}=\sum_{k=1}^{n}\frac{\partial f_{l}}{\partial p_{k}}dp_{k}+\frac{\partial f_{l}}{\partial t}dt=0,\quad l=1,2,\cdot\cdot\cdot,m.\label{constraint01}
\end{equation}

If we consider $dt=0$ and replace $dp_{k}$ by the corresponding
\textit{virtual displacement }$\delta p_{k}$ we can rewrite expression
(\ref{constraint01}) as:

\begin{equation}
\sum_{k=1}^{n}a_{lk}\delta p_{k}=0,\quad l=1,2,\cdot\cdot\cdot,m,\label{constraint02}
\end{equation}
where $a_{lk}=\partial f_{l}/\partial p_{k}.$ Expression (\ref{constraint02})
implies a dependence between the virtual displacements\textit{\ }$\delta p_{k}.$
In order to reduce the number of virtual displacements to only independent
ones we can use Lagrange multipliers $\lambda_{1},\lambda_{2},\cdot\cdot\cdot,\lambda_{m}$.
So, we can put together the equations (\ref{constraint02}) using
the expression:

\begin{equation}
\int_{t_{1}}^{t_{2}}\sum_{k=1}^{n}\sum_{l=1}^{m}\lambda_{l}a_{lk}\delta p_{k}=0.\label{constraint03}
\end{equation}

Therefore, by assuming that the Hamilton's principle holds for holonomic
systems we can incorporate expression (\ref{constraint03}) in the
variational technique used to get Lagrange equations (\ref{Lagran01})
and to obtain:

\begin{equation}
\int_{t_{1}}^{t_{2}}dt\sum_{k=1}^{n}\left(\frac{\partial L}{\partial p_{k}}-\frac{d}{dt}\left(\frac{\partial L}{\partial\overset{\cdot}{p_{k}}}\right)+\sum_{l=1}^{m}\lambda_{l}a_{lk}\right)\delta p_{k}=0.\label{constraint04}
\end{equation}

We shall remember that the virtual displacements $\delta q_{k}$ are
connected by the $m$ equations (\ref{constraint02}). Besides, the
Lagrange multipliers $\lambda_{1},\lambda_{2},\cdot\cdot\cdot,\lambda_{m}$
remains at our disposal. So, let us suppose that we can choose these
multipliers such that:

\begin{equation}
\frac{\partial L}{\partial p_{k}}-\frac{d}{dt}\left(\frac{\partial L}{\partial\overset{\cdot}{p_{k}}}\right)+\sum_{l=1}^{m}\lambda_{l}a_{lk}=0,\quad k=n-m+1,\cdot\cdot\cdot,m.\label{constraint05}
\end{equation}

By substituting this expression in the integral (\ref{constraint04})
we render:

\begin{equation}
\int_{t_{1}}^{t_{2}}dt\sum_{k=1}^{n-m}\left(\frac{\partial L}{\partial p_{k}}-\frac{d}{dt}\left(\frac{\partial L}{\partial\overset{\cdot}{p_{k}}}\right)+\sum_{l=1}^{m}\lambda_{l}a_{lk}\right)\delta p_{k}=0.\label{constraint06}
\end{equation}

Once we have $m$ constraint equations in (\ref{constraint02}) the
only virtual displacements $\delta p_{k}$ involved in expression
(\ref{constraint06}) are the independent ones. Therefore, it follows
that:

\begin{equation}
\frac{\partial L}{\partial p_{k}}-\frac{d}{dt}\left(\frac{\partial L}{\partial\overset{\cdot}{p_{k}}}\right)+\sum_{l=1}^{m}\lambda_{l}a_{lk}=0,\quad k=1,2,\cdot\cdot\cdot,n-m.\label{constraint07}
\end{equation}

Expressions (\ref{constraint05}) and (\ref{constraint07}) give the
complete set of Lagrange's equations for holonomic systems. However,
the expressions involves $n+m$ unknowns, namely the $n$ coordinates
$p_{k}$ and the $m$ multipliers $\lambda_{l}.$ So, we must add
to the final result the constraints give by expression (\ref{constraint01}).

Therefore, by putting together expressions (\ref{constraint05}),
(\ref{constraint07}) and (\ref{constraint00}) we find that the desired
solution must satisfies the equations:

\begin{equation}
\frac{\partial L}{\partial p_{i}}-\frac{d}{dt}\left(\frac{\partial L}{\partial\overset{\cdot}{p_{i}}}\right)+\sum_{l=1}^{m}\lambda_{l}a_{lk}=0,\quad k=1,2,\cdot\cdot\cdot,n,\label{constraint09-a}
\end{equation}

\begin{equation}
f_{l}(p_{1},p_{2},\cdot\cdot\cdot,p_{n},t)=0,,\quad l=1,2,\cdot\cdot\cdot,m.\label{constraint09-b}
\end{equation}

\section{D-NURBS Formulation \label{sec:DNUBS-Formulation}}

The idea is to submit an initial NURBS curve, given by expression
(\ref{nurbs01}) to a Newtonian dynamics generated by an external
potential, internal (elastic) and dissipation forces. Therefore, a
natural way to parameterize the evolution of the curve along the time
is: 
\begin{equation}
\mathbf{c}(u,t)=\frac{\sum_{i=0}^{n}\mathbf{p}_{i}(t)w_{i}(t)B_{i,k}\left(u\right)}{\sum_{i=0}^{n}w_{i}(t)B_{i,k}\left(u\right)}\label{eq:NURBS1DTime}
\end{equation}

So, the control points $\mathbf{p}_{i}(t)$ and the weights $w_{i}(t)$
becomes time-dependent while the rational functions (\ref{rational-spline00})
remain u-dependent only. Therefore, the control points and the weights
become the degrees of freedom of the system evolution; and so, they
compose the generalized coordinates which are concatenated as follows
\cite{terzopoulos_dynamic_1994}: 
\begin{equation}
\mathbf{p}\left(t\right)=\left[\left(\mathbf{p}_{0}^{T},w_{0}\right)\quad\left(\mathbf{p}_{1}^{T},w_{1}\right)\quad\ldots\quad\left(\mathbf{p}_{n}^{T},w_{n}\right)\right]^{T}\in\Re^{4\left(n+1\right)},\label{CoordGeneralizada00}
\end{equation}
where we have the control points vector and the weights vector specified,
respectively, by:

\begin{equation}
\mathbf{p}_{b}\left(t\right)=\left[\mathbf{p}_{0}^{T}\quad\mathbf{p}_{1}^{T}\quad\ldots\quad\mathbf{p}_{n}^{T}\right]^{T}\in\Re^{3\left(n+1\right)},\label{CoordGeneralizada01}
\end{equation}
\begin{equation}
\mathbf{p}_{w}\left(t\right)=\left[w_{0}\quad w_{1}\quad\ldots\quad w_{n}\right]^{T}\in\Re^{\left(n+1\right)}.\label{CoordGeneralizada02}
\end{equation}

A fundamental element in the D-NURBS development is the associated
Jacobian, defined as follows:

\begin{equation}
J=\left[\begin{array}{cccc}
\left[B_{0}\left(u,\mathbf{\mathbf{p}}\right),\frac{\partial\mathbf{c}}{\partial w_{0}}\right], & \left[B_{1}\left(u,\mathbf{p}\right),\frac{\partial\mathbf{c}}{\partial w_{1}}\right], & \cdots & \left[B_{n}\left(u,\mathbf{p}\right),\frac{\partial\mathbf{c}}{\partial w_{n}}\right]\end{array}\right]\in\Re^{3\times4\left(n+1\right)}\label{Jacobiana00}
\end{equation}
where:

\begin{equation}
B_{i}\left(u,\mathbf{p}\right)=\left[\begin{array}{ccc}
\frac{\partial\mathbf{c}}{\partial p_{ix}} & \frac{\partial\mathbf{c}}{\partial p_{iy}} & \frac{\partial\mathbf{c}}{\partial p_{iz}}\end{array}\right]=\left[\begin{array}{ccc}
N_{i,k}\left(u,\mathbf{p}_{w}\right) & 0 & 0\\
0 & N_{i,k}\left(u,\mathbf{p}_{w}\right) & 0\\
0 & 0 & N_{i,k}\left(u,\mathbf{p}_{w}\right)
\end{array}\right]\label{Jacobiana01}
\end{equation}
with (see expression \ref{rational-spline00}): 
\begin{equation}
N_{i,k}\left(u,\mathbf{p}_{w}\right)=\frac{w_{i}(t)B_{i,k}\left(u\right)}{\sum_{j=0}^{n}w_{j}(t)B_{j,k}\left(u\right)},\label{Jacobiana02}
\end{equation}
and: 
\begin{equation}
\frac{\partial\mathbf{c}}{\partial w_{i}}=\frac{\sum_{j=0}^{n}(\mathbf{p}_{i}(t)-\mathbf{p}_{j}(t))w_{j}B_{i,k}\left(u\right)B_{j,k}\left(u\right)}{\left(\sum_{j=0}^{n}w_{j}(t)B_{j,k}\left(u\right)\right)^{2}}.\label{Jacobiana03}
\end{equation}

We shall observe that $B_{i}\left(u,\mathbf{p}\right)\in\Re^{3\times3}$
and $\frac{\partial c}{\partial w_{i}}\in\Re^{3}$, com $i=0,\ldots,n$
and consequently $J\in\Re^{3\times4\left(n+1\right)}$. We can concatenate
the $B_{i^{\prime}s}$ and $\frac{\partial c}{\partial w_{is}}$ according
to the following matrices: 
\begin{equation}
B=\left[B_{0}(u,\mathbf{p}),B_{1}(u,\mathbf{p}),\cdot\cdot\cdot,B_{n}(u,\mathbf{p})\right]\in\Re^{3\times3\left(n+1\right)}\label{eq:B-Matrix}
\end{equation}
\begin{equation}
W=\left[\frac{\partial\mathbf{c}}{\partial w_{0}},\frac{\partial\mathbf{c}}{\partial w_{1}},\cdot\cdot\cdot,\frac{\partial\mathbf{c}}{\partial w_{n}}\right]\in\Re^{3\times\left(n+1\right)}\label{eq:W-Matrix}
\end{equation}

The advantages of defining the matrices $J$, $B$, $W$ and the vectors
$\mathbf{p}_{b}$ and $\mathbf{p}_{w}$ becomes clear by observing
that:

\begin{eqnarray*}
J\mathbf{p}=\left[\begin{array}{cccc}
\left[B_{0}\left(u,\mathbf{p}\right),\frac{\partial\mathbf{c}}{\partial w_{0}}\right], & \left[B_{1}\left(u,\mathbf{p}\right),\frac{\partial\mathbf{c}}{\partial w_{1}}\right], & \cdots & ,\left[B_{n}\left(u,\mathbf{p}\right),\frac{\partial\mathbf{c}}{\partial w_{n}}\right]\end{array}\right]\left[\begin{array}{c}
p_{0x}\\
p_{0y}\\
p_{0z}\\
w_{0}\\
p_{1x}\\
p_{1y}\\
p_{1z}\\
w_{1}\\
\cdot\cdot\cdot\\
\cdot\cdot\cdot\\
p_{nx}\\
p_{ny}\\
p_{nz}\\
w_{n}
\end{array}\right]\\
=\\
=\frac{\sum_{i=0}^{n}w_{i}(t)B_{i,k}(u)\mathbf{p}_{i}(t)}{\sum_{j=0}^{n}w_{j}(t)B_{j,k}(u)}+\sum_{i=0}^{n}\left(\frac{\sum_{j=0}^{n}(\mathbf{p}_{i}(t)-\mathbf{p}_{j}(t))w_{j}(t)B_{i,k}(u)B_{j,k}(u)}{\left(\sum_{j=0}^{n}w_{j}(t)B_{j,k}(u)\right)^{2}}\right)w_{i}(t)=
\end{eqnarray*}
\begin{equation}
B\mathbf{p}_{b}+W\mathbf{p}_{w}.
\end{equation}

But, with a simple algebra we can show that:

\begin{equation}
W\mathbf{p}_{w}=0.\label{propriedade00}
\end{equation}
.

Therefore,

\begin{equation}
J\mathbf{p}=B\mathbf{p}_{b}.\label{propriedade00-1}
\end{equation}

However, by remembering expression (\ref{eq:NURBS1DTime}) it is clear
that:

\begin{equation}
\mathbf{c}(u,t)=B\mathbf{p}_{b}.\label{eq:NURBS1D2}
\end{equation}

Henceforth, from expressions (\ref{propriedade00-1}) and (\ref{eq:NURBS1D2})
we get that:

\begin{equation}
\mathbf{c}(u,\mathbf{p})=J\mathbf{p}.\label{propriedade02}
\end{equation}

Other important properties that can be easily proved are:

\begin{equation}
\frac{dJ}{dt}\cdot\mathbf{p}\left(t\right)=0.\label{propriedade01}
\end{equation}

\begin{equation}
\frac{d\mathbf{c}(u,\mathbf{p})}{dt}=J\cdot\frac{d\mathbf{p}}{dt}.\label{propriedade03}
\end{equation}

The next step is to compute the kinetic and (generalized) potential
terms to be inserted in the Lagrangian given by expression (\ref{Lagrangeana00}).

\subsection{Kinetic Energy $T$\label{subsec:Kinetic-Energy}}

In this work we focus on the D-NURBS formulation for a continuous
parametric curve subject to a force field. So, we shall consider a
(constant) linear mass density distribution $\mu$. Therefore, the
kinetic energy is computed by: 
\begin{equation}
T=\frac{1}{2}\int_{u}\mu\left\Vert \frac{d\mathbf{c}}{dt}\right\Vert ^{2}du,\label{cinetica00}
\end{equation}
where $\frac{d\mathbf{c}}{dt}$ is the curve velocity. By applying
expression~(\ref{propriedade03}) we observe that:

\begin{equation}
\left\Vert \frac{d\mathbf{c}}{dt}\right\Vert ^{2}=\left(J\dot{\mathbf{p}}\right)^{T}.\left(J\dot{\mathbf{p}}\right).\label{cinetica01-1}
\end{equation}

So, if we insert expression (\ref{cinetica01-1}) into kinetic energy
(\ref{cinetica00}) we obtain: 
\begin{equation}
T=\frac{1}{2}\int_{u}\mu\left(J\dot{\mathbf{p}}\right)^{T}.\left(J\dot{\mathbf{p}}\right)du.\label{cinetica01}
\end{equation}
which becomes: 
\begin{equation}
T=\frac{1}{2}\int_{u}\mu\dot{\mathbf{p}}^{T}J^{T}J\dot{\mathbf{p}}du,\label{cinetica02}
\end{equation}

Once $\dot{\mathbf{p}}$ does not depend on the parameter $u$, we
can rewrite expression (\ref{cinetica02}) as: 
\begin{equation}
T=\frac{1}{2}\dot{\mathbf{p}}^{T}M\dot{\mathbf{p}},\label{cinetica03}
\end{equation}
where:

\begin{equation}
M=M(\mathbf{p})=\int_{u}\mu J^{T}Jdu\in\Re^{4\left(n+1\right)\times4\left(n+1\right)},\label{cinetica04-1}
\end{equation}
is called the mass matrix.

\subsection{Energy Dissipation $F$ \label{subsec:Dissipation}}

Formally, the idea is to consider a velocity-dependent potential $F$
such that, when introduced in the Euler-Lagrange equations (\ref{Geral-Euler-Lagrange-04})
generates a velocity-dependent dissipative force. In order to perform
this task let us suppose that $F$ satisfies: 
\begin{equation}
\frac{dF}{dt}=-\frac{1}{2}\int_{u}\gamma\left\Vert \frac{d\mathbf{c}}{dt}\right\Vert ^{2}du\quad\Rightarrow\quad F\left(t\right)=-\frac{1}{2}\int_{t=0}^{t}\int_{u}\gamma\left\Vert \frac{d\mathbf{c}}{dt}\right\Vert ^{2}dudt,\quad\label{dissip00}
\end{equation}
where the constant $\gamma$ is the the damping density. By performing
an analogous development of section \ref{subsec:Kinetic-Energy} we
obtain:

\begin{equation}
\frac{dF}{dt}=-\frac{1}{2}\dot{\mathbf{p}}^{T}D\dot{\mathbf{p}},\label{dissip03}
\end{equation}
where $D\in\Re^{4\left(n+1\right)\times4\left(n+1\right)}$, the damping
matrix, is computed by: 
\begin{equation}
D=D(\mathbf{p})=\int_{u}\gamma J^{T}Jdu.\label{dissip03-1}
\end{equation}

\subsection{Potential for Conservative Forces \label{subsec:Potential-Consev}}

The internal and external conservative forces are introduced in the
D-NURBS Lagrangian through the potentials $E_{int}$ and $E_{ext}$,
respectively. We compute the former by using the \emph{thin-plate
}model \cite{terzopoulos_deformable_1988}: 
\begin{equation}
E_{int}\left(\mathbf{p}\right)=\frac{1}{2}\int_{u}\left(\alpha\left\Vert \frac{d\mathbf{c}}{du}\right\Vert ^{2}+\beta\left\Vert \frac{d^{2}\mathbf{c}}{du^{2}}\right\Vert ^{2}\right)du,\label{Eint00}
\end{equation}
where $\alpha$ is the elasticity and $\beta$ the rigidity parameter
of the curve. Using the expression (\ref{propriedade02}) and the
fact that the generalize coordinates vector $\mathbf{p}$ does not
depends on the parameter $u$ (see expression (\ref{CoordGeneralizada00}))
we can show that:

\begin{equation}
\frac{d\mathbf{c}}{du}=\frac{d}{du}\left(J\mathbf{p}\right)=J_{u}\mathbf{p}.\label{eq:derivada-u}
\end{equation}

Obviously the same is true for the second derivative respect to the
parameter $u$ . Therefore: 
\begin{equation}
E_{int}\left(\mathbf{p}\right)=\frac{1}{2}\int_{u}\left(\alpha\left(J_{u}\mathbf{p}\right)^{T}\left(J_{u}\mathbf{p}\right)+\beta\left(J_{uu}\mathbf{p}\right)^{T}\left(J_{uu}\mathbf{p}\right)\right)du.\label{Eint01}
\end{equation}

Once $\left(J_{u}\mathbf{p}\right)^{T}\left(J_{u}\mathbf{p}\right)=\mathbf{p}^{T}J_{u}^{T}J_{u}\mathbf{p}$
it follows: 
\begin{equation}
E_{int}\left(\mathbf{p}\right)=\frac{1}{2}\int_{u}\left(\alpha\mathbf{p}^{T}J_{u}^{T}J_{u}\mathbf{p}+\beta\mathbf{p}^{T}J_{uu}^{T}J_{uu}\mathbf{p}\right)du,\label{Eint02}
\end{equation}
and, consequently:

\begin{equation}
E_{int}\left(\mathbf{p}\right)=\frac{1}{2}\mathbf{p}^{T}K\mathbf{p},\label{Eint03}
\end{equation}
where the matrix $K=K(\mathbf{p})\in\Re^{4\left(n+1\right)\times4\left(n+1\right)}$,
named the stiffness matrix, is given by:

\begin{equation}
K(\mathbf{p})=\int_{u}\left(\alpha J_{u}^{T}J_{u}+\beta J_{uu}^{T}J_{uu}\right)du.\label{Eint04}
\end{equation}

The external potential $E_{ext}$ generates the force fields, like
gravity, that act on the system. According to expression (\ref{Geral-Euler-Lagrange-04}),
they are computed by the gradient of the potential $E_{ext}$ respect
to the generalized coordinates:

\begin{equation}
\frac{\partial E_{ext}}{\partial\mathbf{p}}=\frac{1}{2}\left(\frac{\partial E_{ext}}{\partial p_{0x}},\frac{\partial E_{ext}}{\partial p_{0y}},\frac{\partial E_{ext}}{\partial p_{0z}},\frac{\partial E_{ext}}{\partial w_{0}};\frac{\partial E_{ext}}{\partial p_{1x}},\frac{\partial E_{ext}}{\partial p_{1y}},\frac{\partial E_{ext}}{\partial p_{1z}},\frac{\partial E_{ext}}{\partial w_{1}};\cdot\cdot\cdot;\frac{\partial E_{ext}}{\partial p_{nx}},\frac{\partial E_{ext}}{\partial p_{ny}},\frac{\partial E_{ext}}{\partial p_{nz}},\frac{\partial E_{ext}}{\partial w_{n}}\right)^{T}.\label{Eext00}
\end{equation}

\subsection{Euler-Lagrange Equations for D-NURBS \label{subsec:DNURBS-Equations}}

Now, we insert the kinetic energy and potentials just computed in
the Euler-Lagrange equations given by expression (\ref{Geral-Euler-Lagrange-04}).
Besides, we must observe that the matrices $M$, $D$ and $K$ are
all symmetric and for a quadratic form $g=\mathbf{p}^{T}A\mathbf{p}$
with $A$ symmetric we have $\frac{\partial g}{\partial\mathbf{p}}=2A\mathbf{p}.$
Therefore:

\bigskip{}

\begin{itemize}
\item $\frac{d}{dt}\left(\frac{\partial T}{\partial\overset{\cdot}{\mathbf{p}}}\right)=\frac{d}{dt}\left(\frac{\partial}{\partial\overset{\cdot}{\mathbf{p}}}\left(\frac{1}{2}\dot{\mathbf{p}}^{T}M\dot{\mathbf{p}}\right)\right)=\frac{d}{dt}\left(\frac{1}{2}2M\dot{\mathbf{p}}\right)=\frac{d}{dt}\left(M\dot{\mathbf{p}}\right)=M\ddot{\mathbf{p}}+\dot{M}\dot{\mathbf{p}}$\label{term0}
\item $\frac{d}{dt}\left(\frac{\partial F}{\partial\overset{\cdot}{\mathbf{p}}}\right)=\frac{\partial}{\partial\overset{\cdot}{\mathbf{p}}}\left(\frac{dF}{dt}\right)=\frac{\partial}{\partial\overset{\cdot}{\mathbf{p}}}\left(-\frac{1}{2}\dot{\mathbf{p}}^{T}D\dot{\mathbf{p}}\right)=-\frac{1}{2}2D\dot{\mathbf{p}}=-D\dot{\mathbf{p}}$\label{term1}
\item $\frac{\partial T}{\partial\mathbf{p}}=\frac{\partial}{\partial\mathbf{p}}\left(\frac{1}{2}\dot{\mathbf{p}}^{T}M\dot{\mathbf{p}}\right)=\frac{1}{2}\left(\overset{\cdot}{\mathbf{p}}\right)^{T}\frac{\partial M}{\partial\mathbf{p}}\overset{\cdot}{\mathbf{p}}$\label{term2}
\item $\frac{\partial F}{\partial\mathbf{p}}=-\frac{1}{2}\int_{t=0}^{t}\left(\int_{u}\dot{\mathbf{p}}^{T}\frac{\partial D}{\partial\mathbf{p}}\dot{\mathbf{p}}du\right)dt$
\item $\frac{\partial E_{int}}{\partial\mathbf{p}}=\frac{\partial}{\partial\mathbf{p}}\left(\frac{1}{2}\mathbf{p}^{T}K\mathbf{p}\right)=\frac{1}{2}2K\mathbf{p}+\frac{1}{2}\left[\mathbf{p}^{T}\frac{\partial K}{\partial\mathbf{p}}\mathbf{p}\right]^{T}=K\mathbf{p}+\frac{1}{2}\left[\mathbf{p}^{T}\frac{\partial K}{\partial\mathbf{p}}\mathbf{p}\right]^{T}$\label{term4} 
\end{itemize}
By substituting these expressions in the Euler-Lagrange equation:

\begin{equation}
\frac{d}{dt}\left(\frac{\partial T}{\partial\overset{\cdot}{\mathbf{p}}}-\frac{\partial F}{\partial\overset{\cdot}{\mathbf{p}}}\right)+\left(\frac{\partial E_{int}}{\partial\mathbf{p}}\right)=-\frac{\partial E_{ext}}{\partial\mathbf{p}}+\left(\frac{\partial T}{\partial\mathbf{p}}-\frac{\partial F}{\partial\mathbf{p}}\right),\label{Euler-Lagrange-04}
\end{equation}
we get: 
\begin{equation}
\left(M\overset{\cdot\cdot}{\mathbf{p}}+\overset{\cdot}{M}\overset{\cdot}{\mathbf{p}}\right)-\left(-D\overset{\cdot}{\mathbf{p}}\right)+K\mathbf{p}+\left[\frac{1}{2}\mathbf{p}^{T}\frac{\partial K}{\partial\mathbf{p}}\mathbf{p}\right]^{T}=-\frac{\partial E_{ext}}{\partial\mathbf{p}}+\left[\frac{1}{2}\left(\overset{\cdot}{\mathbf{p}}\right)^{T}\frac{\partial M}{\partial\mathbf{p}}\overset{\cdot}{\mathbf{p}}\right]^{T}+\frac{1}{2}\int_{t=0}^{t}\left(\int_{u}\dot{\mathbf{p}}^{T}\frac{\partial D}{\partial\mathbf{p}}\dot{\mathbf{p}}du\right)dt,\label{Euler-Lagrange-04-1}
\end{equation}
which can be rewritten as follows by just re-arranging the terms:
\begin{equation}
M\overset{\cdot\cdot}{\mathbf{p}}+D\overset{\cdot}{\mathbf{p}}+K\mathbf{p}=-\frac{\partial E_{ext}}{\partial\mathbf{p}}+\left[\frac{1}{2}\left(\overset{\cdot}{\mathbf{p}}\right)^{T}\frac{\partial M}{\partial\mathbf{p}}\overset{\cdot}{\mathbf{p}}\right]^{T}-\overset{\cdot}{M}\overset{\cdot}{\mathbf{p}}-\left[\frac{1}{2}\mathbf{p}^{T}\frac{\partial K}{\partial\mathbf{p}}\mathbf{p}\right]^{T}+\frac{1}{2}\int_{t=0}^{t}\left(\int_{u}\dot{\mathbf{p}}^{T}\frac{\partial D}{\partial\mathbf{p}}\dot{\mathbf{p}}du\right)dt.\label{Euler-Lagrange-04-2}
\end{equation}
However, in the Appendices A and B we shown that: 
\begin{equation}
I\dot{\mathbf{p}}=\dot{M}\overset{\cdot}{\mathbf{p}}-\frac{1}{2}\left(\overset{\cdot}{\mathbf{p}}\right)^{T}\frac{\partial M}{\partial\mathbf{p}}\overset{\cdot}{\mathbf{p}},\label{simplificando}
\end{equation}

\begin{equation}
\left[\frac{1}{2}\mathbf{p}^{T}\frac{\partial K}{\partial\mathbf{p}}\mathbf{p}\right]=0,\label{appendixB00}
\end{equation}
where: 
\begin{equation}
I=\int\mu J^{T}\overset{\cdot}{J}du.\label{eq:operador-I}
\end{equation}
Therefore, if we neglect the effects of the last integral term, we
can finally write the governing equation for D-NURBS as: 
\begin{equation}
M\overset{\cdot\cdot}{\mathbf{p}}+D\overset{\cdot}{\mathbf{p}}+K\mathbf{p}=-\frac{\partial E_{ext}}{\partial\mathbf{p}}-I\overset{\cdot}{\mathbf{p}},\label{Evolution-Eq00}
\end{equation}
where the matrix $I$ is computed by equation (\ref{eq:operador-I}).

\subsection{External Forces \label{subsec:External-Force}}

In this section we consider an external potential which in cartesian
coordinates has the general form:

\begin{equation}
E_{ext}=\int_{u}P\left(x,y,z\right)du
\end{equation}
where $P\left(x,y,z\right)$ is a potential density function.

So, 
\begin{equation}
E_{ext}\left(\mathbf{p}\right)=\int_{u}P\left(x\left(\mathbf{p},u\right),y\left(\mathbf{p},u\right),z\left(\mathbf{p},u\right)\right)du
\end{equation}

In cartesian coordinates, the external force is given by:

\begin{equation}
F_{ext=}\int_{u}\left[\begin{array}{c}
\frac{\partial P}{\partial x}\\
\frac{\partial P}{\partial y}\\
\frac{\partial P}{\partial z}
\end{array}\right]du.
\end{equation}

However, we must write the external force respect to the generalized
coordinates, following the expression (\ref{Eext00}). For instance,
let us consider the term: 
\begin{equation}
\frac{\partial E_{ext}}{\partial p_{0x}}=\int_{u}\left[\frac{\partial}{\partial p_{0x}}P\left(x\left(\mathbf{p},u\right),y\left(\mathbf{p},u\right),z\left(\mathbf{p},u\right)\right)\right]du
\end{equation}

From the Chain-Rule: 
\begin{equation}
\frac{\partial E_{ext}}{\partial p_{0x}}=\int_{u}\left[\frac{\partial P}{\partial x}\frac{\partial x}{\partial p_{0x}}+\frac{\partial P}{\partial y}\frac{\partial y}{\partial p_{0x}}+\frac{\partial P}{\partial z}\frac{\partial z}{\partial p_{0x}}\right]du.
\end{equation}

But, from the expression (\ref{eq:NURBS1DTime}) we observe that:
\begin{equation}
\frac{\partial y}{\partial p_{0x}}=\frac{\partial z}{\partial p_{0x}}=0.
\end{equation}

Therefore: 
\begin{equation}
\frac{\partial E_{ext}}{\partial p_{0x}}=\int_{u}\left[\frac{\partial P}{\partial x}\frac{\partial x}{\partial p_{0x}}\right]du
\end{equation}

Analogously we can find:

\begin{equation}
\frac{\partial E_{ext}}{\partial p_{0y}}=\int_{u}\left[\frac{\partial P}{\partial y}\frac{\partial y}{\partial p_{0y}}\right]du,\quad\frac{\partial E_{ext}}{\partial p_{0z}}=\int_{u}\left[\frac{\partial P}{\partial z}\frac{\partial z}{\partial p_{0z}}\right]du.
\end{equation}

On the other hand: 
\begin{equation}
\frac{\partial E_{ext}}{\partial w_{0}}=\int_{u}\left[\frac{\partial}{\partial w_{0}}P\left(x\left(\mathbf{p},u\right),y\left(\mathbf{p},u\right),z\left(\mathbf{p},u\right)\right)\right]du,
\end{equation}

and so:{\small 
\begin{equation}
\frac{\partial E_{ext}}{\partial w_{0}}=\int_{u}\left[\frac{\partial P}{\partial x}\frac{\partial x}{\partial w_{0}}+\frac{\partial P}{\partial y}\frac{\partial y}{\partial w_{0}}+\frac{\partial P}{\partial z}\frac{\partial z}{\partial w_{0}}\right]du=\int_{u}\left[\frac{\partial x}{\partial w_{0}},\frac{\partial y}{\partial w_{0}},\frac{\partial z}{\partial w_{0}}\right]\left[\frac{\partial P}{\partial x},\frac{\partial P}{\partial y},\frac{\partial P}{\partial z}\right]^{T}du
\end{equation}
}{\small \par}

So, by using expression (\ref{Jacobiana01}), we find that the above
results can be grouped in the following matricial expression: 
\begin{equation}
\left[\begin{array}{c}
\frac{\partial E_{ext}}{\partial p_{0x}}\\
\frac{\partial E_{ext}}{\partial p_{0y}}\\
\frac{\partial E_{ext}}{\partial p_{0z}}\\
\frac{\partial E_{ext}}{\partial w_{0}}
\end{array}\right]=\int_{u}\left[\begin{array}{ccc}
N_{0,k} & 0 & 0\\
0 & N_{0,k} & 0\\
0 & 0 & N_{0,k}\\
\frac{\partial x}{\partial w_{0}} & \frac{\partial y}{\partial w_{0}} & \frac{\partial z}{\partial w_{0}}
\end{array}\right]\cdot\left[\begin{array}{c}
\frac{\partial P}{\partial x}\\
\frac{\partial P}{\partial y}\\
\frac{\partial P}{\partial z}
\end{array}\right]du=\int_{u}\left[\begin{array}{c}
N_{0,k}\frac{\partial P}{\partial x}\\
N_{0,k}\frac{\partial P}{\partial y}\\
N_{0,k}\frac{\partial P}{\partial z}\\
\frac{\partial x}{\partial w_{0}}\frac{\partial P}{\partial x}+\frac{\partial y}{\partial w_{0}}\frac{\partial P}{\partial y}+\frac{\partial z}{\partial w_{0}}\frac{\partial P}{\partial z}
\end{array}\right]du
\end{equation}

Generalizing for $i=0..n$ we have: 
\begin{equation}
\frac{\partial E_{ext}}{\partial p}=\int_{u}\left[\begin{array}{c}
\begin{array}{ccc}
N_{0,k}\\
 & N_{0,k}\\
 &  & N_{0,k}\\
\frac{\partial x}{\partial w_{0}} & \frac{\partial y}{\partial w_{0}} & \frac{\partial z}{\partial w_{0}}\\
N_{1,k}\\
 & N_{1,k}\\
 &  & N_{1,k}\\
\frac{\partial x}{\partial w_{1}} & \frac{\partial y}{\partial w_{1}} & \frac{\partial z}{\partial w_{1}}\\
\cdot & \cdot & \cdot\\
N_{n,k}\\
 & N_{n,k}\\
 &  & N_{n,k}\\
\frac{\partial x}{\partial w_{n}} & \frac{\partial y}{\partial w_{n}} & \frac{\partial z}{\partial w_{n}}
\end{array}\end{array}\right]\cdot\left[\begin{array}{c}
\frac{\partial P}{\partial x}\\
\frac{\partial P}{\partial y}\\
\frac{\partial P}{\partial z}
\end{array}\right]du=\int_{u}J^{T}f\left(x,y,z\right)du,
\end{equation}
where $f\left(x,y,z\right)$ is the external force field density defined
by the gradient of the potential density $P$ in cartesian coordinates
$\left(x,y,z\right).$ Therefore, the external force field in the
generalized coordinates is given by:

\begin{equation}
f_{\mathbf{p}}\left(\mathbf{p}\right)=\int_{u}J^{T}f\left(x\left(\mathbf{p},u\right),y\left(\mathbf{p},u\right),z\left(\mathbf{p},u\right)\right)du.\label{external-force-field}
\end{equation}

Particularly, in the case of the gravitational potential for a particle
we have: 
\begin{equation}
E=-mgy,\label{particle-grav-potential}
\end{equation}
where $m$ is the mass particle, $g$ is the gravitational field intensity
and $y$ gives the particle position (its height) in the vertical
axes. For a $1D$ continuous system, a curve in the two-dimensional
Euclidean space, the gravitational potential can be computed by a
generalization of expression (\ref{particle-grav-potential}) given
by: 
\begin{equation}
E_{ext}=-\int_{u}\mu gydu
\end{equation}
where $\mu$ is the linear mass density (constant), like before. Therefore,
the potential density is: 
\begin{equation}
P\left(x,y\right)=-\mu gy.
\end{equation}
and the force field density is given by:

\begin{equation}
\nabla P\left(x,y\right)=-\mu g(\frac{\partial y}{\partial x},\frac{\partial y}{\partial y})^{T}=-\mu g(0,1)^{T}
\end{equation}

So, according to expression (\ref{external-force-field}), the external
force field $f_{\mathbf{p}}$ is: 
\begin{equation}
f_{\mathbf{p}}\left(\mathbf{p}\right)=-\int_{u}\mu J^{T}g\left(\begin{array}{c}
0\\
1
\end{array}\right)du=-\int_{u}\mu J^{T}\left(\begin{array}{c}
0\\
g
\end{array}\right)du.
\end{equation}

\subsection{D-NURBS with Constraints \label{subsec:DNURBS-Constraints}}

\bigskip{}
 In the case of linear constraints, equations (\ref{constraint00})
become:

\begin{equation}
A\mathbf{p}+\mathbf{d}=0,\label{linear-contraint00}
\end{equation}
where $A\in\Re^{m\times n},$ with $m<n$, is a constant matrix and
$\mathbf{d}\in\Re^{m}$ is a constant vector. In this case, we can
choose a set of $n-m,$ say $\mathbf{q}=\left(q_{1},q_{2},\cdot\cdot\cdot,q_{n-m}\right)$
independent variables and explicitly write the $m$ remaining ones
as a function of the $\mathbf{q}$ vector, which will be the new generalized
coordinates. In fact, if we write equation (\ref{linear-contraint00})
in the form:

\begin{equation}
\left(\begin{array}{cccc}
a_{11} & a_{12} & \cdot\cdot\cdot & a_{1m}\\
a_{21} & a_{22} & \cdot\cdot\cdot & a_{2m}\\
 &  & \cdot\cdot\cdot\\
a_{m1} & a_{m2} & \cdot\cdot\cdot & a_{mm}
\end{array}\right)\left(\begin{array}{c}
p_{1}\\
p_{2}\\
\cdot\cdot\cdot\\
p_{m}
\end{array}\right)+\left(\begin{array}{cccc}
a_{1,m+1} & a_{1,m+2} & \cdot\cdot\cdot & a_{1,n}\\
a_{2,m+1} & a_{2,m+2} & \cdot\cdot\cdot & a_{2,n}\\
 &  & \cdot\cdot\cdot\\
a_{m,m+1} & a_{m,m+2} & \cdot\cdot\cdot & a_{m,n}
\end{array}\right)\left(\begin{array}{c}
p_{m+1}\\
p_{m+2}\\
\cdot\cdot\cdot\\
p_{n}
\end{array}\right)=-\mathbf{d,}\label{linear-contraint01}
\end{equation}
or simply:

\[
G_{1}\overline{\mathbf{q}}+G_{2}\mathbf{q=-d}
\]
where \ $G_{1}\in\Re^{m\times m}$ and $G_{2}\in\Re^{m\times\left(n-m\right)}$
are the first and second matrices of expression (\ref{linear-contraint01})
and $\overline{\mathbf{q}}=\left(p_{1},p_{2},\cdot\cdot\cdot,p_{m}\right)^{T},$
$\mathbf{q=}\left(p_{m+1},p_{m+2},\cdot\cdot\cdot,p_{n}\right)^{T}.$
Then, by supposing that $p_{1},\cdot\cdot\cdot,p_{m}$ can be choosen
such that $G_{1}$ is non-singular, we have:

\begin{equation}
\overline{\mathbf{q}}\mathbf{=-}G_{1}^{-1}G_{2}\mathbf{q}-G_{1}^{-1}\mathbf{d}.\label{linear-contraint02}
\end{equation}

Let $I\in\Re^{\left(n-m\right)\times\left(n-m\right)}$ and from the
observation that $G_{1}^{-1}G_{2}\in\Re^{m\times\left(n-m\right)}$
and using expression (\ref{linear-contraint02}) it is clear that
the matrix:

\begin{equation}
G=\left[\begin{array}{c}
\mathbf{-}G_{1}^{-1}G_{2}\\
I
\end{array}\right]\in\Re^{n\times\left(n-m\right)},\label{linear-contraint03}
\end{equation}
allows to write:

\begin{equation}
\mathbf{p}=G\mathbf{\mathbf{q+}d}_{0}\label{linear-contraint04}
\end{equation}
with $\mathbf{d}_{0}=\left[G_{1}^{-1}\mathbf{d},\mathbf{0}\right]^{T}\in\Re^{n\times1}$

The equations (\ref{constraint09-a}) can be written in compact form
as:

\begin{equation}
A^{T}\mathbf{\lambda}=-\left(M\overset{\cdot\cdot}{\mathbf{p}}+D\overset{\cdot}{\mathbf{p}}+K\mathbf{p+}\frac{\partial E_{ext}}{\partial\mathbf{p}}+I\overset{\cdot}{\mathbf{p}}\right).\label{linear-contraint05}
\end{equation}

So, using expressions (\ref{linear-contraint04}) we can observe that:

\begin{equation}
\overset{\cdot}{\mathbf{p}}=G\overset{\cdot}{\mathbf{q}}\mathbf{,\quad}\overset{\cdot\cdot}{\mathbf{p}}=G\overset{\cdot\cdot}{\mathbf{q}}.\label{linear-contraint06}
\end{equation}

Therefore, by substituting expressions (\ref{linear-contraint04})
and (\ref{linear-contraint06}) in equation (\ref{linear-contraint05})
and using the fac that $A=\left[G_{1}\quad G_{2}\right]$ \ we obtain:

\[
\left[\begin{array}{c}
G_{1}^{T}\\
G_{2}^{T}
\end{array}\right]\mathbf{\lambda=}-\left(MG\overset{\cdot\cdot}{\mathbf{q}}+DG\overset{\cdot}{\mathbf{q}}+K\left(G\mathbf{\mathbf{q+}d}_{0}\right)\mathbf{+}\frac{\partial E_{ext}}{\partial\mathbf{p}}\frac{\partial\mathbf{p}}{\partial\mathbf{q}}+IG\overset{\cdot}{\mathbf{q}}\right),
\]

If we multiply both sides by $G^{T}$, where the matrix $G$ is defined
in expression (\ref{linear-contraint03}) we obtain:

\[
\left[\mathbf{-}\left(G_{1}^{-1}G_{2}\right)^{T}\quad I\right]\times\left[\begin{array}{c}
G_{1}^{T}\\
G_{2}^{T}
\end{array}\right]\mathbf{\lambda=}-G^{T}\times\left(MG\overset{\cdot\cdot}{\mathbf{q}}+DG\overset{\cdot}{\mathbf{q}}+K\left(G\mathbf{\mathbf{q+}d}_{0}\right)\mathbf{+}\frac{\partial E_{ext}}{\partial\mathbf{p}}+IG\overset{\cdot}{\mathbf{q}}\right),
\]

\[
0=-G^{T}MG\overset{\cdot\cdot}{\mathbf{q}}-G^{T}DG\overset{\cdot}{\mathbf{q}}-G^{T}K\left(G\mathbf{\mathbf{q+}d}_{0}\right)-G^{T}\frac{\partial E_{ext}}{\partial\mathbf{p}}-G^{T}IG\overset{\cdot}{\mathbf{q}}
\]

\[
G^{T}MG\overset{\cdot\cdot}{\mathbf{q}}+G^{T}DG\overset{\cdot}{\mathbf{q}}+G^{T}KG\mathbf{\mathbf{q}}=-G^{T}\frac{\partial E_{ext}}{\partial\mathbf{p}}-G^{T}IG\overset{\cdot}{\mathbf{q}}-G^{T}K\mathbf{d}_{0}.
\]

If we name:

\[
M_{q}=G^{T}MG;\:\: D_{q}=G^{T}DG;\:\: K_{q}=G^{T}KG;\:\: f_{q}=-G^{T}\frac{\partial E_{ext}}{\partial\mathbf{p}};\:\: I_{q}=G^{T}IG,
\]
then, we get the D-NURBS evolution equation subject to the linear
constraints given by:

\begin{equation}
M_{q}\overset{\cdot\cdot}{\mathbf{q}}+D_{q}\overset{\cdot}{\mathbf{q}}+K_{q}\mathbf{\mathbf{q}}=f_{q}-I_{q}\overset{\cdot}{\mathbf{q}}-G^{T}K\mathbf{d}_{0}.\label{Euler-Lagrange-Constrained}
\end{equation}

\subsection{Numerical Implementation \label{subsec:Numerical}}

The equation (\ref{Evolution-Eq00}), as well as its constrained counterpart
in expresson (\ref{Euler-Lagrange-Constrained}), does not have in
general analytical solution and so we have to use a numerical approach
to solve it with the desired precision. The equation (\ref{Evolution-Eq00})
is a second order ordinary differential equation. Besides, it is important
to observe that the matrices $M$, $D$, $K$ depends on the integration
of products of the rational B-spline functions (\ref{rational-spline00})
and their derivatives of first and second order respect to the variable
$u$.

Therefore, the numerical solution of expression (\ref{Evolution-Eq00})
can be performed by finite difference methods (FDM) in time. Besides,
we need a numerical scheme for computing the integrals, as described
next.

\subsubsection{Matrices Computation \label{subsubsec:MatComp}}

The matrices $M$, $D$, $K$ that appears in D-NURBS evolution equation
are given by expressions (\ref{cinetica04-1}),(\ref{dissip03-1}),
and (\ref{Eint04}), respectively. They involve derivatives of zero,
first and second order of $J$ respect to the variable $u$. For instance,
for matrix $K=\left(k_{ij}\right)\in\Re^{4\left(n+1\right)\times4\left(n+1\right)}$
we have:

\begin{equation}
k_{ij}=\int_{u}f_{ij}\left(u\right)du={\displaystyle \sum_{i=0}^{n-1}}\int_{u_{i}}^{u_{i+1}}f_{ij}\left(u\right)du,\label{eq:computation-K}
\end{equation}
where:

\begin{equation}
f_{ij}=\alpha\left(\frac{\partial\mathbf{j}_{i}}{\partial u}\right)^{T}\left(\frac{\partial\mathbf{j}_{j}}{\partial u}\right)+\beta\left(\frac{\partial^{2}\mathbf{j}_{i}}{\partial u^{2}}\right)^{T}\left(\frac{\partial^{2}\mathbf{j}_{j}}{\partial u^{2}}\right),
\end{equation}
with $\mathbf{j}_{i}$ means the collum $i$ of the Jacobian $J$.

The computation of each term in the summation in expression (\ref{eq:computation-K})
can be performed by Gauss quadrature \cite{chapra_numerical_2009}.
An analogous scheme can be used to compute the other matrices.

In our implementation we have developed a numerical approach based
on isogeometric analysis following the recipe of \cite{cottrell_isogeometric_2009}.
Our implementation avoids the cost of assembling the global matrices
$M$, $D$ and $K$. For this, we calculate the matrices of each element
individually, where the elements are constructed by partitioning the
knots vector. Figure \ref{fig:BuildingElementIGA} shows building
elements $\left(e1,e2,e3,e4\right)$ from open knots vector $\mathbf{v}=\left(0,0,0,0.25,0.50,0.75,1,1,1\right)$.
Here we will have six control points $\left(n-k+1=8-3+1=6\right)$
which according \cite{cottrell_isogeometric_2009} will be distributed
over the elements by following expression 
\begin{equation}
\mathbf{E}=\left\{ e1=\left(p_{1},p_{2},p_{3}\right),e2=\left(p_{2},p_{3},p_{4}\right),e3=\left(p_{3},p_{4},p_{5}\right),e4=\left(p_{4},p_{5},p_{6}\right)\right\} 
\end{equation}

\begin{center}
\begin{figure}[h]
\begin{centering}
\includegraphics[scale=0.4]{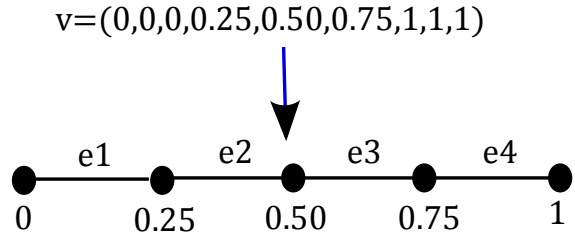} 
\par\end{centering}

\caption{Building elements from knots vector $\mathbf{v}=\left(0,0,0,0.25,0.50,0.75,1,1,1\right)$}

\label{fig:BuildingElementIGA} 
\end{figure}

\par\end{center}

Generally a open knots vector can be partitioned into $n_{e}$ elements
expressed by

\begin{equation}
n_{e}=\left|\mathbf{E}\right|=n-2(k-1)
\end{equation}

\subsubsection{Numerical Scheme for Time Integration \label{subsubsec:TimeIntegration}}

Let the D-NURBS evolution equation: 
\begin{equation}
M\overset{\cdot\cdot}{\mathbf{p}}+D\overset{\cdot}{\mathbf{p}}+K\mathbf{p}=f_{\mathbf{p}}(\mathbf{p})-I\dot{\mathbf{p}}\label{eq:MotionSemRestricao01}
\end{equation}
where $f_{\mathbf{p}}=\int J^{T}f(x,y,z)du$ and $I(p)=\int\mu J^{T}\dot{J}du$,
according to Appendix A and sections \ref{subsec:DNURBS-Equations}-\ref{subsec:External-Force}.

Let us consider the following numerical scheme: 
\begin{equation}
\ddot{\mathbf{p}}=\frac{\mathbf{p}^{\left(t+\triangle t\right)}-2\mathbf{p}^{(t)}+\mathbf{p}^{(t-\triangle t)}}{\left(\triangle t\right)^{2}}\label{eq:discretization-second}
\end{equation}
\begin{equation}
\dot{\mathbf{p}}=\frac{\mathbf{p}^{\left(t+\triangle t\right)}-\mathbf{p}^{(t-\triangle t)}}{2\triangle t}\label{eq:discretization-first}
\end{equation}

If we substitute these expressions in equation (\ref{eq:MotionSemRestricao01})
we obtain: 
\begin{equation}
M\left(\frac{\mathbf{p}{}^{\left(t+\triangle t\right)}-2\mathbf{p}{}^{(t)}+\mathbf{p}^{(t-\triangle t)}}{\left(\triangle t\right)^{2}}\right)+D\left(\frac{\mathbf{p}{}^{\left(t+\triangle t\right)}-\mathbf{p}^{\left(t-\triangle t\right)}}{2\triangle t}\right)+K\mathbf{p}{}^{\left(t+\triangle t\right)}=f_{\mathbf{p}}-I\dot{\left(\frac{\mathbf{p}{}^{\left(t+\triangle t\right)}-\mathbf{p}{}^{\left(t-\triangle t\right)}}{2\triangle t}\right)}\label{eq:MotionSemRestricao02}
\end{equation}
where $M$, $D$, $K$ and $I$ are supposed to be computed at time
$t+\triangle t.$ We shall be careful about the term $I\dot{\left(\frac{\mathbf{p}{}^{\left(t+\triangle t\right)}-\mathbf{p}{}^{\left(t-\triangle t\right)}}{2\triangle t}\right)}$.
Following its definition in expression (\ref{eq:compute-I}) and expression
(\ref{eq:discretization-first}) we can write:

\begin{equation}
I\dot{\mathbf{p}}=\int\mu\left(J^{T}\right)^{(t+\triangle t)}\left(\dot{J}\right)^{(t+\triangle t)}du\left(\frac{\mathbf{p}{}^{\left(t+\triangle t\right)}-\mathbf{p}{}^{\left(t-\triangle t\right)}}{2\triangle t}\right)\label{eq:Ip_ponto01}
\end{equation}

Therefore, we can rewrite equation~(\ref{eq:Ip_ponto01}) as: 
\begin{align}
\notag\\
I\dot{\mathbf{p}}=\frac{1}{2\triangle t}\int\mu\left(J^{T}\right)^{(t+\triangle t)}\left[\left(\dot{J}\right)^{(t+\triangle t)}\mathbf{p}{}^{\left(t+\triangle t\right)}-\left(\dot{J}\right)^{(t+\triangle t)}\mathbf{p}{}^{\left(t-\triangle t\right)}\right]du & .\label{eq:equation-Ipo}
\end{align}

By using the fact that $\dot{J}^{(t+\triangle t)}\mathbf{p}^{(t+\triangle t)}=0$
we simplify expression (\ref{eq:equation-Ipo}) to:

\begin{align*}
I\dot{\mathbf{p}}=-\frac{1}{2\triangle t}\int\mu\left(J^{T}\right)^{(t+\triangle t)}\left[\left(\dot{J}\right)^{(t+\triangle t)}\mathbf{p}{}^{\left(t-\triangle t\right)}\right]du & .
\end{align*}

Using the approximation:

\begin{align*}
\left(\dot{J}\right)^{(t+\triangle t)}=\left(\frac{J{}^{\left(t+\triangle t\right)}-J{}^{\left(t-\triangle t\right)}}{2\triangle t}\right),
\end{align*}

we get:

\begin{align}
I\dot{\mathbf{p}}=-\frac{1}{2\triangle t}\int\mu\left(J^{T}\right)^{(t+\triangle t)}\left[\left(\frac{J{}^{\left(t+\triangle t\right)}-J{}^{\left(t-\triangle t\right)}}{2\triangle t}\right)\mathbf{p}{}^{\left(t-\triangle t\right)}\right]du\notag\\
=-\frac{1}{4\left(\triangle t\right)^{2}}\int\mu\left(J^{T}\right)^{(t+\triangle t)}\left[J{}^{\left(t+\triangle t\right)}-J{}^{\left(t-\triangle t\right)}\right]\mathbf{p}{}^{\left(t-\triangle t\right)}du\notag\\
=-\frac{1}{4\left(\triangle t\right)^{2}}\int\mu\left(J^{T}\right)^{(t+\triangle t)}\left[J{}^{\left(t+\triangle t\right)}\mathbf{p}{}^{\left(t-\triangle t\right)}-J{}^{\left(t-\triangle t\right)}\mathbf{p}{}^{\left(t-\triangle t\right)}\right]du\notag\\
=-\frac{1}{4\left(\triangle t\right)^{2}}\left(\int\mu\left(J^{T}\right)^{(t+\triangle t)}J{}^{\left(t+\triangle t\right)}\mathbf{p}{}^{\left(t-\triangle t\right)}-\int\mu\left(J^{T}\right)^{(t+\triangle t)}J{}^{\left(t-\triangle t\right)}\mathbf{p}{}^{\left(t-\triangle t\right)}\right)du\notag\\
=-\frac{1}{4\left(\triangle t\right)^{2}}\left(M{}^{\left(t+\triangle t\right)}\mathbf{p}{}^{\left(t-\triangle t\right)}-\int\mu\left(J^{T}\right)^{(t+\triangle t)}\mathbf{c}{}^{\left(t-\triangle t\right)}du\right)\label{eq:Ip_ponto02}
\end{align}

So, by substituting expression (\ref{eq:Ip_ponto02}) in (\ref{eq:MotionSemRestricao02})
it renders:

\[
M\left(\frac{p{}^{\left(t+\triangle t\right)}-2p{}^{(t)}+p^{(t-\triangle t)}}{\left(\triangle t\right)^{2}}\right)+D\left(\frac{p{}^{\left(t+\triangle t\right)}-p^{\left(t-\triangle t\right)}}{2\triangle t}\right)+Kp{}^{\left(t+\triangle t\right)}=
\]
\begin{equation}
f_{p}+\frac{1}{4\left(\triangle t\right)^{2}}\left(M{}^{\left(t+\triangle t\right)}p{}^{\left(t-\triangle t\right)}-\int\mu\left(J^{T}\right)^{(t+\triangle t)}c{}^{\left(t-\triangle t\right)}du\right),\label{eq:MotionSemRestricao03}
\end{equation}

If we multiply both sides of expression (\ref{eq:MotionSemRestricao02})
to $\times4\left(\triangle t\right)^{2}$ and rearrange the terms
we get:

\begin{equation}
\left(4M+2\triangle tD+4\left(\triangle t\right)^{2}K\right)\mathbf{p}{}^{\left(t+\triangle t\right)}=4\left(\triangle t\right)^{2}f_{\mathbf{p}}+8M\mathbf{p}{}^{(t)}-\left(3M-2\triangle tD\right)\mathbf{p}^{(t-\triangle t)}-\int\mu J^{T}\mathbf{c}{}^{\left(t-\triangle t\right)}du\label{eq:MotionSemRestricao05}
\end{equation}

This expression can be written as:

\begin{equation}
A_{0}^{(t+\triangle t)}\mathbf{p}{}^{\left(t+\triangle t\right)}=A_{1}^{(t,t-\triangle t)},\label{DNURBS-iterative-scheme}
\end{equation}
where:

\begin{equation}
A_{0}^{(t+\triangle t)}=4M+2\triangle tD+4\left(\triangle t\right)^{2}K,
\end{equation}
and,

\begin{equation}
A_{1}^{(t,t-\triangle t)}=4\left(\triangle t\right)^{2}f_{p}+8M\mathbf{p}{}^{(t)}-\left(3M-2\triangle tD\right)\mathbf{p}^{(t-\triangle t)}-\int\mu J^{T}\mathbf{c}{}^{\left(t-\triangle t\right)}du.
\end{equation}

Therefore, once initial conditions $\mathbf{p}(0)=\mathbf{p}_{0}$
and $\dot{\mathbf{p}}(0)=\mathbf{v}_{0}$ are given, we can use the
approximation: 
\begin{equation}
\frac{\mathbf{p}(0)-\mathbf{p}(0-\triangle t)}{\triangle t}=\mathbf{v}_{0}.
\end{equation}
to write: 
\begin{equation}
\mathbf{p}(-\triangle t)=\mathbf{p}(0)-\triangle t\mathbf{v}_{0},
\end{equation}
and, consequently, we can start the iterative scheme given by expression
(\ref{DNURBS-iterative-scheme}).

The complexity for computing the expression (\ref{DNURBS-iterative-scheme})
depends on the algorithm for calculating the matrices $M$, $D$,
$K$ and the method used to solve the linear system. Considering that
$n$ is the number of control points, $n_{e}$ is number of elements,
$k$ is the polynomial order of NURBS basis and $n_{g}$ is the number
of quadrature points, the algorithm implemented to compute the matrices
$M$, $D$, $K$ performs the following steps:
\begin{enumerate}
\item For $e=1\ldots n_{e}$ do

\begin{enumerate}
\item Compute the Jacobian matrix block for element ``$e$'' (complexity
$O(n_{g}*n*k)$).
\item Compute mass matrix block for element ``$e$'' (complexity $O(n_{g}*k^{2})$).
\item Compute damping matrix block for element ``$e$'' (complexity $O(n_{g}*k^{2})$).
\item Compute stiffness matrix block for element ``$e$'' (complexity
$O(n_{g}*k^{2})$). 
\end{enumerate}
\end{enumerate}
Therefore, the asymptotic complexity of the whole algorithm is given
by: 
\begin{equation}
O\left(n_{e}n_{g}\left(O\left(nk\right)+O\left(k^{2}\right)+O\left(k^{2}\right)+O\left(k^{2}\right)\right)\right)=O\left(n_{e}n_{g}nk\right).\label{final-complexity-DNURBS}
\end{equation}

We highlight that the computational cost of the D-NURBS evolution
must also consider the numerical method for solving the linear system
\ref{DNURBS-iterative-scheme}. To compute \ref{DNURBS-iterative-scheme}
we have used conjugate gradient method whose complexity is $O\left(n\right)$.
Hence, we can conclude that the expression \ref{DNURBS-iterative-scheme}
has final computational complexity equal to $O\left(n_{e}n_{g}nk\right)$.

\section{Experimental Results \label{sec:Exper}}

We have developed an experimental environment based on the D-NURBS
approach with constraints. In our setting we consider the case of
an elastic wire with $10m$ length with negligible transverse section
fixed at the ends.

The NURBS curve geometry is instantiated using an open knot vector
$\mathbf{v}=\left(0,0,0,0,0.25,0.50,0.75,1,1,1,1\right)$ with basis
functions of order $k=4$ (degree $k-1=3$). Therefore, following
section \ref{sec:NURBS-Rev}, the spline space has dimension $n-k+1=10-4+1=7$,
which means that we have seven controls points. Each point of a NURBS
curve is influenced by $k$ control points. Therefore, to set geometric
constraints that keep the wire fixed at the ends, we must let $k-1$
fixed control points at the ends of the curve. This can be cast in
the linear constraint framework for D-NURBS developed in section \ref{subsec:DNURBS-Constraints}.

Besides, we consider that the wire is subject to a gravitational field
with value $g=9.8m/s^{2}$ and define control points position and
weights at $t=0$ according to table \ref{tab:CondInitialP}. Besides,
we set $\dot{\mathbf{p}}(0)=0$ to complete the initial conditions
for time integration.

\begin{center}
\begin{table}
\begin{centering}
\begin{tabular}{|c|c|c|c|c|}
\hline 
$i$  & $x_{i}$  & $y_{i}$  & $z_{i}$  & $w_{i}$\tabularnewline
\hline 
\hline 
$0$  & -5.00  & 5  & 0  & 1\tabularnewline
\hline 
$1$  & -4.17  & 5  & 0  & 1\tabularnewline
\hline 
$2$  & -2.50  & 5  & 0  & 1\tabularnewline
\hline 
$3$  & 0.00  & 5  & 0  & 1\tabularnewline
\hline 
$4$  & 2.50  & 5  & 0  & 1\tabularnewline
\hline 
$5$  & 4.17  & 5  & 0  & 1\tabularnewline
\hline 
$6$  & 5.00  & 5  & 0  & 1\tabularnewline
\hline 
\end{tabular}
\par\end{centering}

\caption{Initial configuration of the control points and weights (generalized
coordinates) for wire simulation: $\mathbf{p}(0)=(x_{0},y_{0},z_{0},w_{0};x_{1},y_{1},z_{1},w_{1};\cdot\cdot\cdot;x_{6},y_{6},z_{6},w_{6})$. }

\label{tab:CondInitialP} 
\end{table}

\par\end{center}

Figure \ref{fig:SetupDNURBSCable} demonstrates the environment at
time $t=0s$. Here physics parameters were defined as: $\alpha=35$,
$\beta=10$, $\mu=30$, $\gamma=0$. To perform spatial and time integration
we define $10$ points in Gauss quadrature and $\triangle t=0.008s$,
respectively.

\begin{center}
\begin{figure}
\begin{centering}
\includegraphics[scale=0.9]{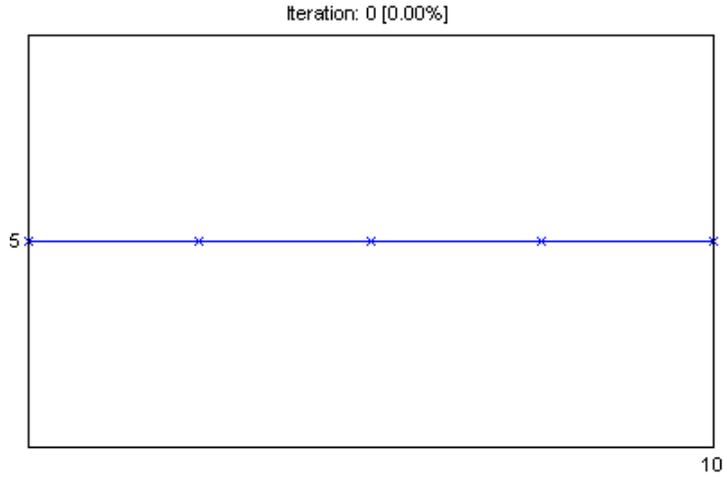} 
\par\end{centering}

\caption{Intial D-NURBS setup for simulation of the elastic wire fixed at the
ends.}

\label{fig:SetupDNURBSCable} 
\end{figure}

\par\end{center}

In our experiments we observed that the evolution of the weights $w_{i}$
may cause unrealistic behaviors and instability, as observed in Figure
\ref{fig:DNURBSUnrestrictW}.(a). As mentioned in \cite{terzopoulos_dynamic_1994},
the weights $w_{i}$ may not have arbitrary finite real values. Negative
values may vanish the denominator of the rational functions in expression
\ref{rational-spline00}. Besides, small weights values may lower
the deformation energy \cite{terzopoulos_dynamic_1994}. Therefore,
some constraint must be included in order to enforce some control
in the weight vector evolution.

In this work we implement this task by a very simple strategy: the
generalized coordinate vector is updated by solving the expression
\ref{Euler-Lagrange-Constrained}, but the weight vector is always
returned to its initial value; that means, $w_{i}=1$ for $i=0,1,\cdot\cdot\cdot,6$,
following Table \ref{tab:CondInitialP}. As observed in Figure \ref{fig:DNURBSUnrestrictW}.(b),
the wire evolution becomes (visually) acceptable in this case.

\begin{center}
\begin{figure}
\subfloat[Wire configuration at iteration $t=1,10,22$ without constrain the
weights evolution.]{\begin{centering}
\includegraphics[scale=0.5]{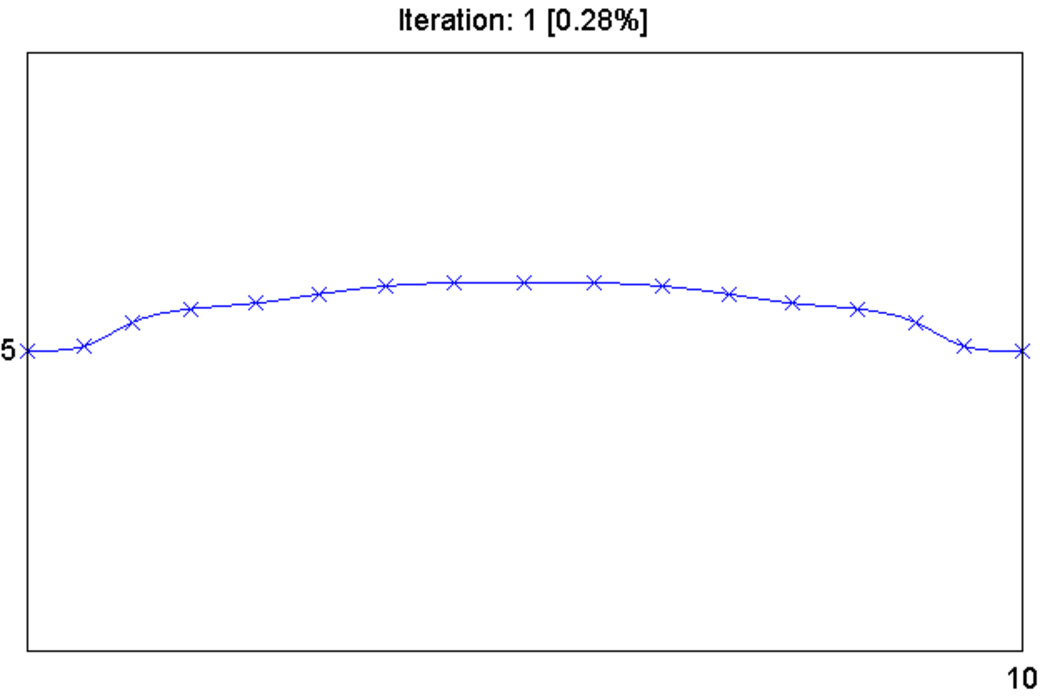}\includegraphics[scale=0.5]{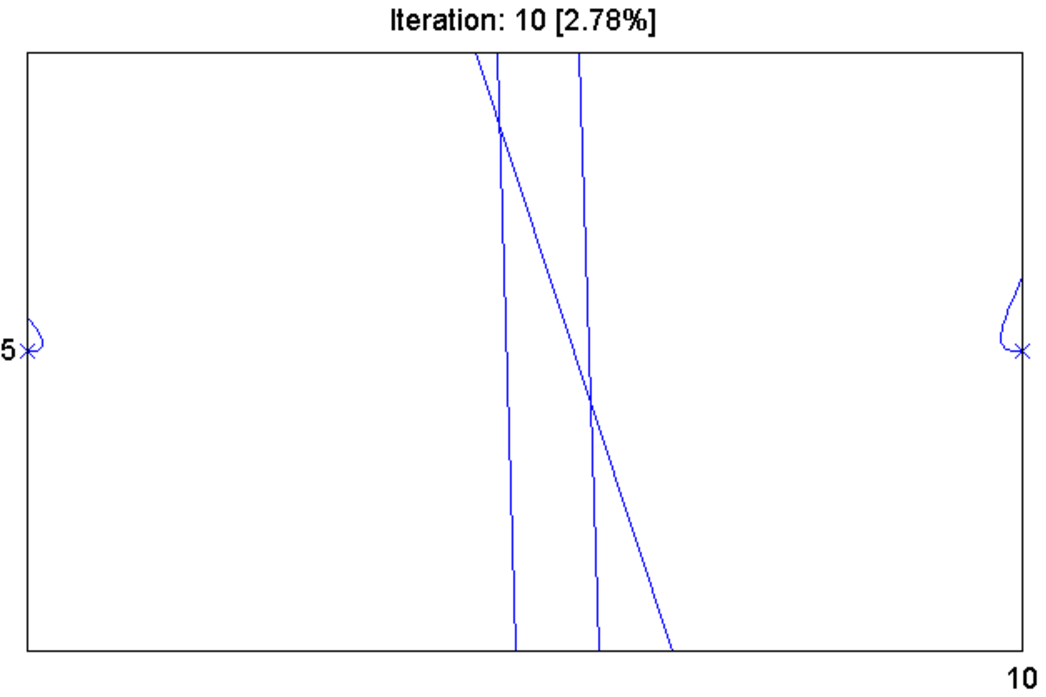}\includegraphics[scale=0.5]{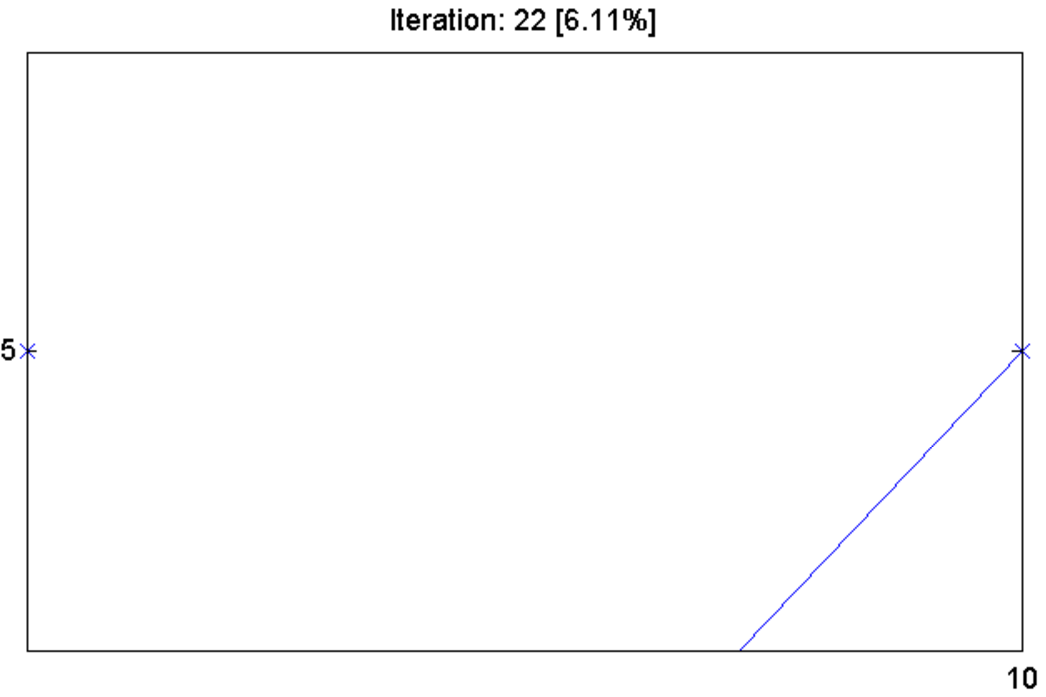} 
\par\end{centering}

}

\medskip{}
 \subfloat[System configuration at iteration $t=1,10,22$ when enforcing weights
$w_{i}=1$ after each iteration.]{\begin{centering}
\includegraphics[scale=0.5]{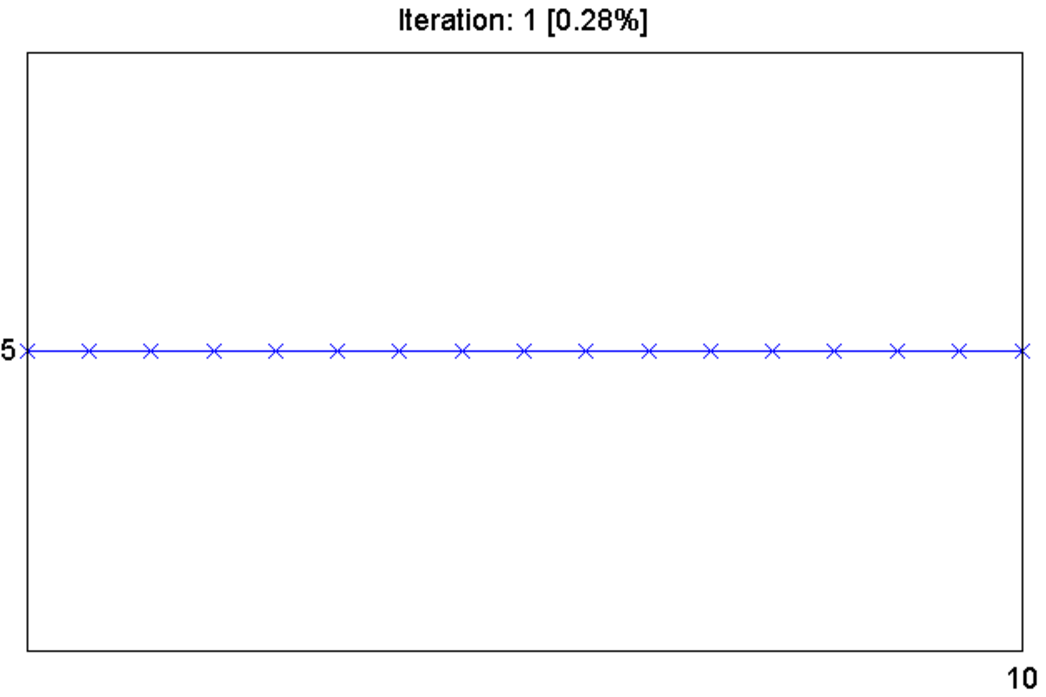}\includegraphics[scale=0.5]{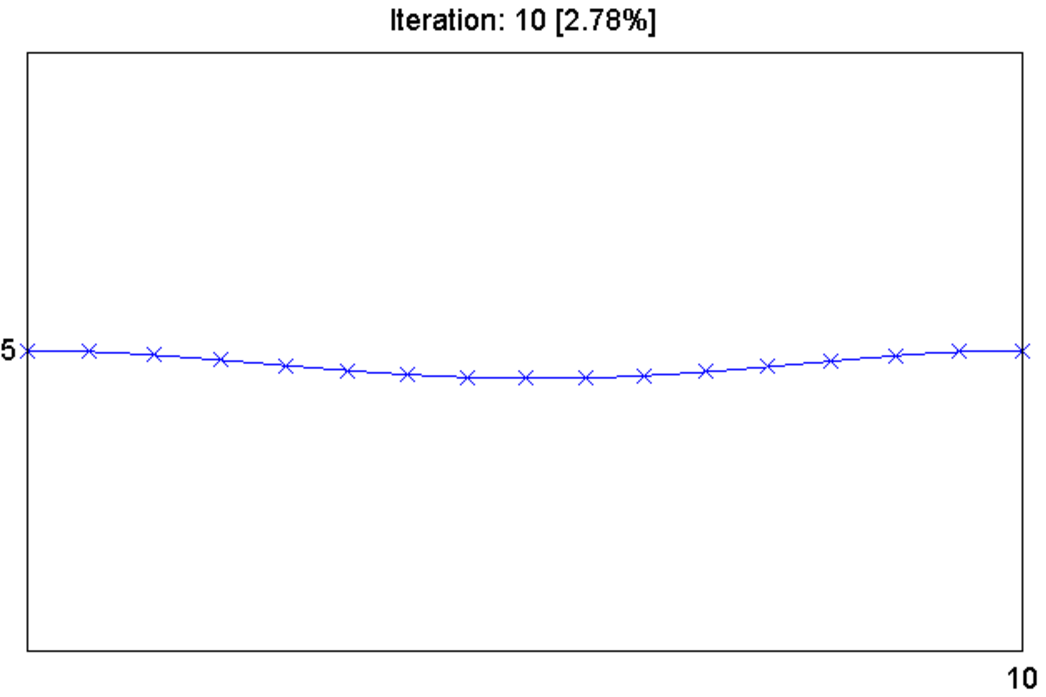}\includegraphics[scale=0.5]{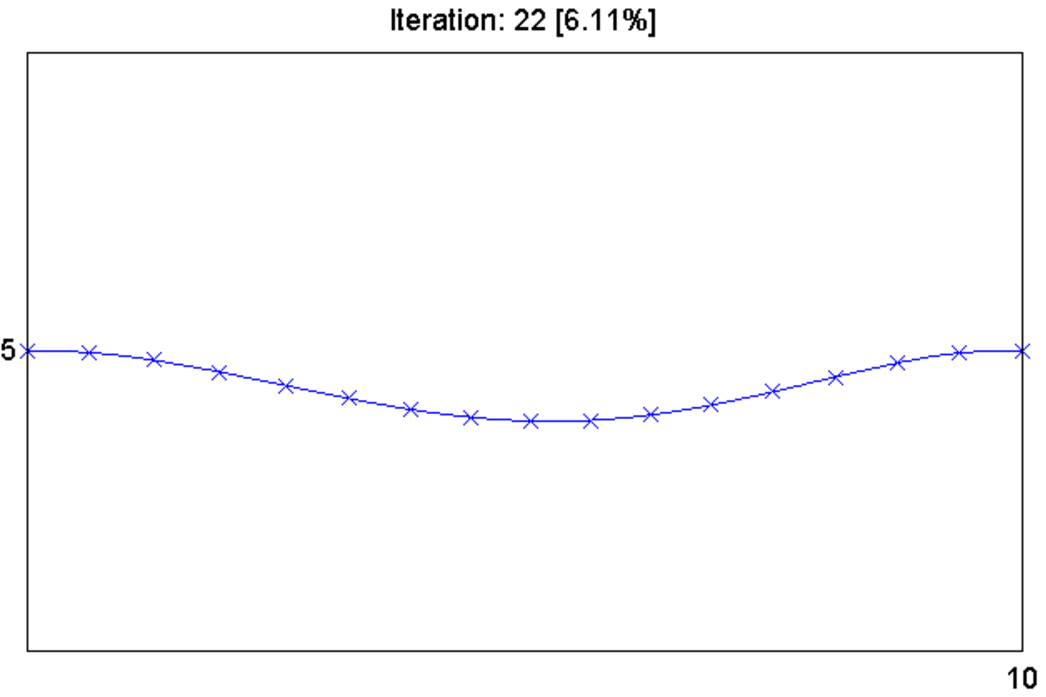} 
\par\end{centering}

}

\caption{D-NURBS behavior for unconstrained and constrained weight vector evolution.}

\label{fig:DNURBSUnrestrictW} 
\end{figure}

\par\end{center}

To study the dynamic evolution of the D-NURBS curve we choose a point
in the center of the wire and followed its amplitude evolution in
time. Figure \ref{fig:DNURBSWithoutDamp} shows its dynamic evolution
without the presence of damping, while Figure \ref{fig:DNURBSWithDamp}
illustrates the dynamic evolution with damping, where $\gamma=5$.
As expected, the former reports a periodic evolution once there are
not dissipative forces and the latter pictures an attenuation of the
amplitude along the time due to the damping.

\begin{center}
\begin{figure}
\begin{centering}
\subfloat[\label{fig:DNURBSWithoutDamp}Amplitude evolution for D-NURBS without
damping.]{\begin{centering}
\includegraphics[scale=0.8]{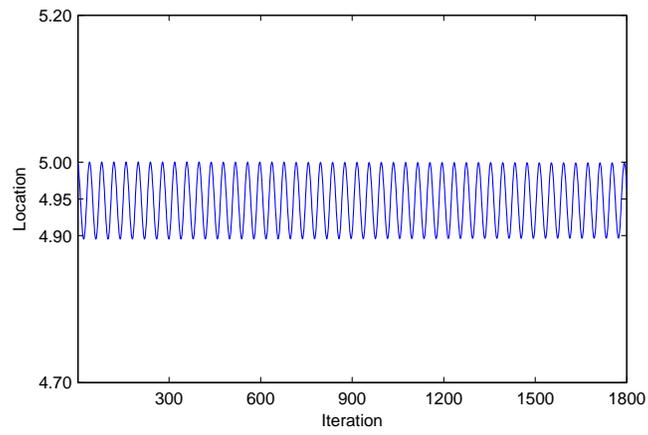} 
\par\end{centering}

}
\par\end{centering}

\begin{centering}
\subfloat[\label{fig:DNURBSWithDamp}Amplitude evolution of D-NURBS with damping.]{\begin{centering}
\includegraphics[scale=0.8]{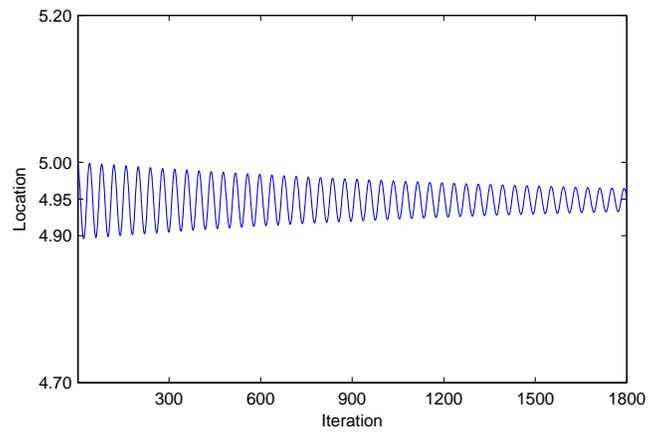} 
\par\end{centering}

}
\par\end{centering}

\caption{Amplitude evolution of elastic wire represented by D-NURBS.}
\end{figure}

\par\end{center}

To analyze the effects of the elasticity and stiffness of D-NURBS
curve we increasing each parameter separately. First, we leave $\beta$
(see equation \ref{Eint00}) with the same value of the initial configuration
and modify $\alpha$. The Figures \ref{fig:DNURBSExAlpha1} and \ref{fig:DNURBSExAlpha2}
show the results.

Similarly, we modify $\beta$ while $\alpha$ remains unchanged. The
results are shown in figures \ref{fig:DNURBSExBeta1} and \ref{fig:DNURBSExBeta2}.
We observe that the system is more sensitive respect to the parameter
$\beta$ than the parameter $\alpha$. In fact, when increasing the
parameter $\beta$ from $11$ to $15$ we observe a drastic change
in the amplitude evolution as highlighted when comparing Figures \ref{fig:DNURBSExBeta1}
and \ref{fig:DNURBSExBeta2}. On the other hand, when changing $\alpha$
from $70$ to $105$ (Figures \ref{fig:DNURBSExAlpha1} and \ref{fig:DNURBSExAlpha2},
respectively) we did not observe a similar behavior.

\begin{center}
\begin{figure}
\begin{centering}
\subfloat[\label{fig:DNURBSExAlpha1} $\alpha=70$ and $\beta=10$.]{\begin{centering}
\includegraphics[scale=0.5]{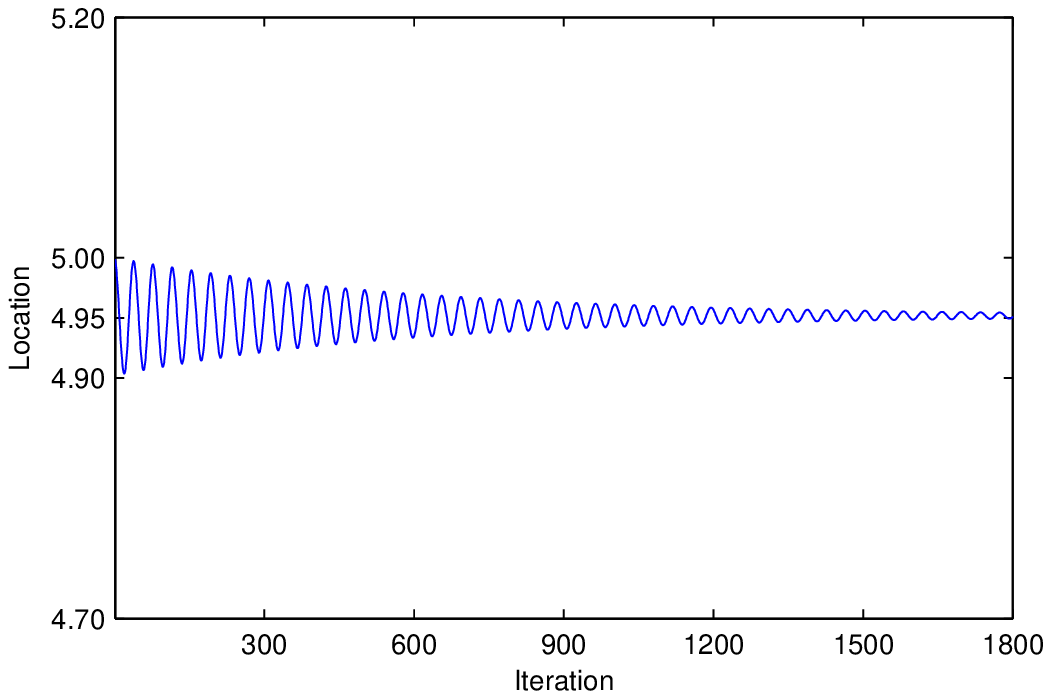} 
\par\end{centering}

}
\par\end{centering}

\begin{centering}
\subfloat[\label{fig:DNURBSExAlpha2} $\alpha=105$ and $\beta=10$.]{\begin{centering}
\includegraphics[scale=0.5]{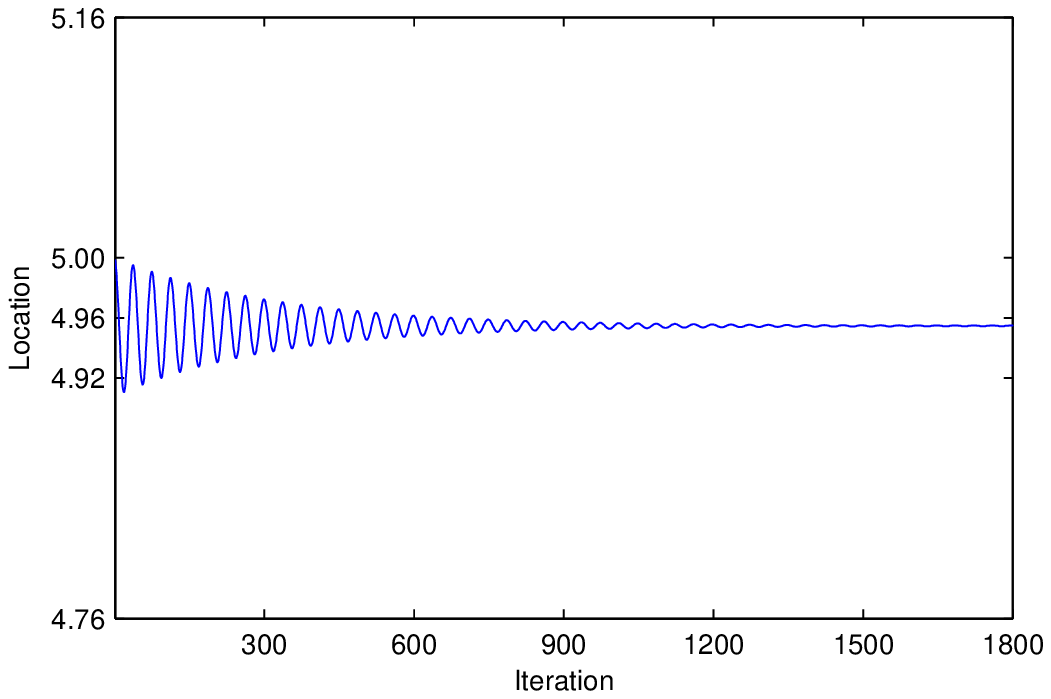} 
\par\end{centering}

}
\par\end{centering}

\begin{centering}
\medskip{}

\par\end{centering}

\begin{centering}
\subfloat[\label{fig:DNURBSExBeta1} $\alpha=35$ and$\beta=11$.]{\begin{centering}
\includegraphics[scale=0.5]{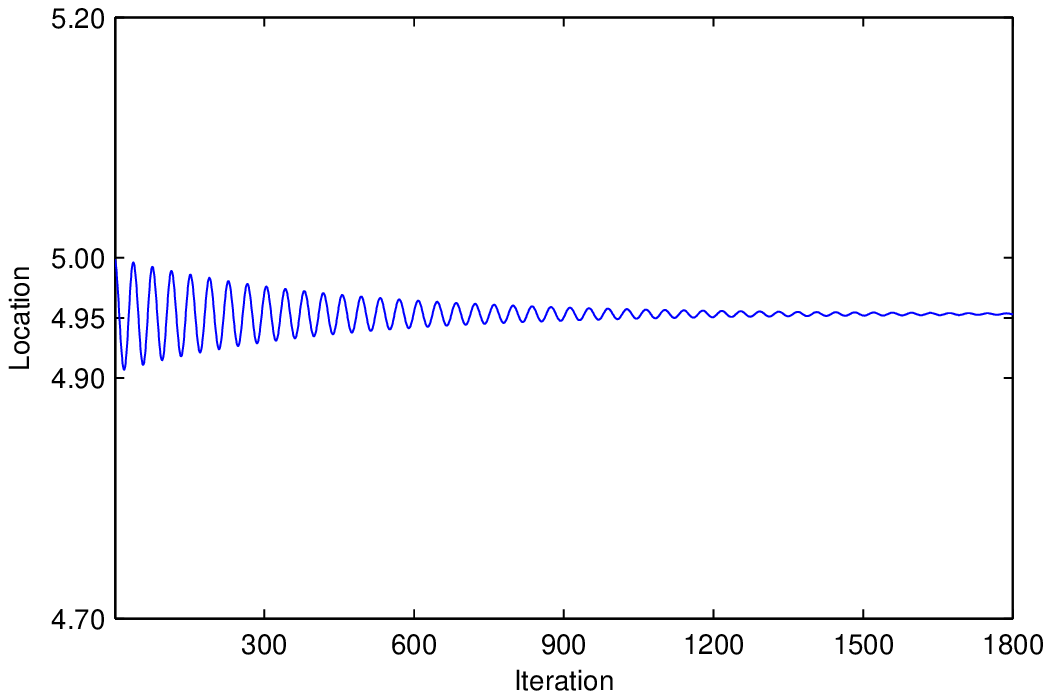} 
\par\end{centering}

}
\par\end{centering}

\begin{centering}
\subfloat[\label{fig:DNURBSExBeta2}$\alpha=35$ and $\beta=15$.]{\begin{centering}
\includegraphics[scale=0.5]{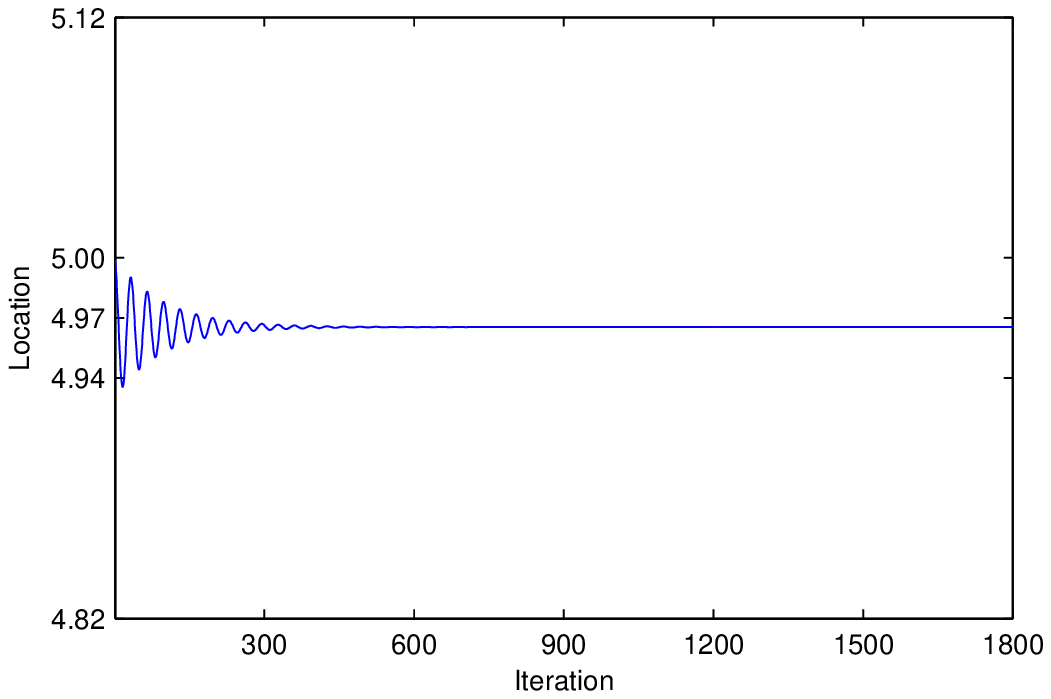} 
\par\end{centering}

}
\par\end{centering}

\caption{Sensitivity of D-NURBS amplitude respect to elasticity $\alpha$ and
stiffness $\beta$.}
\end{figure}

\par\end{center}

The expression (\ref{final-complexity-DNURBS}) shows that the number
of control points has a fundamental role in the computational cost
of the D-NURBS algorithm. Therefore, we perform a runtime analysis
of D-NURBS evolution for different number of controls points. We set
parameters to: $k=3$,$\alpha=35$, $\beta=50$, $\mu=30$, $\gamma=1$
and five points in Gauss quadrature. The host is a Intel Core I5-3210M
at 2.5 Ghz, with 6 GB RAM running a Windows 7 (64bit).

We take $360$ iterations for each configuration and measure the corresponding
CPU time. In order to compare the complexity given by expression (\ref{final-complexity-DNURBS})
and the CPU time for each simulation, we compute the following rates:
\begin{equation}
P^{j}=\frac{T_{{}}^{j+1}-T_{{}}^{j}}{T_{{}}^{j}},\label{porcentagem-cpu-time}
\end{equation}

\begin{equation}
\widehat{P}^{j}=\frac{C_{{}}^{j+1}-C_{{}}^{j}}{C_{{}}^{j}},\label{porcentagem-complexity}
\end{equation}
where $T_{}^{j}$ is the CPU time for configuration $j$ and $C_{}^{j}$
is the asymptotic complexity for the same configuration; that means:

\begin{equation}
C_{}^{j}=360\ast n_{e}^{j}n_{g}^{j}n^{j}k^{j}.
\end{equation}

The configuration ($j=1$) has $20$ control points and $k=4$ and
$360$ D-NURBS iterations are performed$.$ Next, for $j=2$, we increase
the number of control points by $5$, keep $k=4$ and perform $360$
iterations of the algorithm again, and so on. The result is pictured
on Figure \ref{fig:ratetimecomplexity} where the dot blue curve shows
the evolution of expression (\ref{porcentagem-complexity}) and the
red line shows the evolution of expression (\ref{porcentagem-cpu-time}),
both for $j=1,2,\cdot\cdot\cdot,28$ (number of control points $20$,
$25$, $\ldots$, $140$).

\begin{center}
\begin{figure}
\begin{centering}
\includegraphics[scale=0.9]{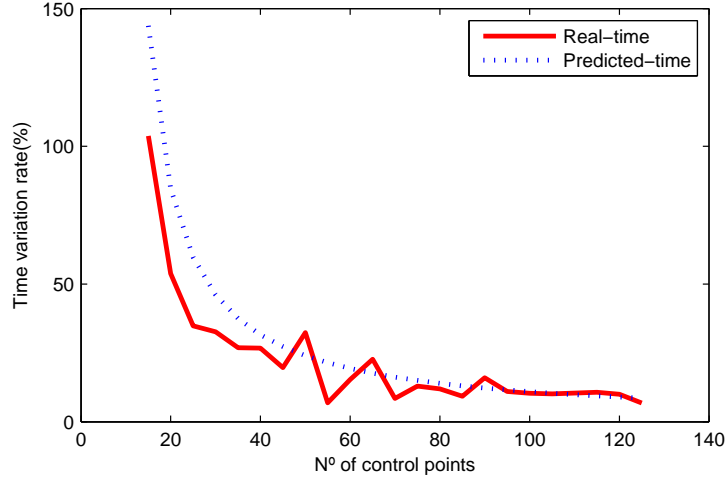} 
\par\end{centering}

\caption{Rates for computational complexity an CPU time given by expressions
(\ref{porcentagem-complexity}) and (\ref{porcentagem-cpu-time}),
respectively. The former is pictured by the blue plot while the latter
by the red one. }

\label{fig:ratetimecomplexity} 
\end{figure}

\par\end{center}

By observing Figure \ref{fig:ratetimecomplexity} we note that as
we increase the control points number we get $P^{j}\rightarrow\widehat{P}^{j}$
. This is the expected behavior for asymptotic function, i.e., for
large $n$ the similarity between real and predicted time becomes
more evident.

\section{Conclusions and Future Works \label{sec:Concl}}

We present a review of D-NURBS approach. We emphasize the formulation
based on the Lagrangian mechanics followed by detailed development
of the governing equations. We used a numerical method based on isogeometric
analysis for the spatial integration used to compute the Jacobian,
mass, damping and stiffness matrix. For validation we performed experiments
with D-NURBS curve and discuss the influence of parameters, effects
of NURBS weights and computational cost. For further works we plan
to evaluate the D-NURBS for 2D and 3D systems.

\bibliographystyle{sbc}
\bibliography{DNURBS}

\begin{thebibliography}{}

\bibitem[Chapra and Canale 2009]{chapra_numerical_2009}
Chapra, S. and Canale, R. (2009).
\newblock {\em Numerical Methods for Engineers}.
\newblock {McGraw-Hill} Education.

\bibitem[Cottrell et~al. 2009]{cottrell_isogeometric_2009}
Cottrell, J., Hughes, T., and Bazilevs, Y. (2009).
\newblock {\em Isogeometric Analysis: Toward Integration of {CAD} and {FEA}}.
\newblock Wiley.

\bibitem[Deusen et~al. 2004]{deusen_elements_2004}
Deusen, O., Ebert, D.~S., Fedkiw, R., Musgrave, F.~K., Prusinkiewicz, P.,
  Roble, D., Stam, J., and Tessendorf, J. (2004).
\newblock The elements of nature: interactive and realistic techniques.
\newblock In {\em {ACM} {SIGGRAPH} 2004 Course Notes}, page~32.

\bibitem[Erleben et~al. 2005]{erleben_physics-based_2005}
Erleben, K., Sporring, J., Henriksen, K., and Dohlman, K. (2005).
\newblock {\em Physics-based Animation (Graphics Series)}.
\newblock Charles River Media, Inc., Rockland, {MA}, {USA}.

\bibitem[Farin 1997]{farin_curves_1997}
Farin, G. (1997).
\newblock {\em Curves and surfaces for computer-aided geometric design: a
  practical guide}.
\newblock Number vol. 1 in Computer science and scientific computing. Academic
  Press.

\bibitem[Goldstein 1981]{goldstein_classical_1981}
Goldstein, H. (1981).
\newblock {\em Classical Mechanics}.
\newblock Addison-Wesley, 2nd edition.

\bibitem[Persiano 1996]{persiano_bases_1996}
Persiano, R.~M. (1996).
\newblock {\em Bases da Modelagem Geometrica}.
\newblock 10a Escola de Computacao.

\bibitem[Piegl and Tiller 1997]{piegl_nurbs_1997}
Piegl, L. and Tiller, L. (1997).
\newblock {\em The Nurbs Book}.
\newblock Monographs in Visual Communication Series. Springer-Verlag {GmbH}.

\bibitem[Qin and Terzopoulos 1996]{qin_d-nurbs:_1996}
Qin, H. and Terzopoulos, D. (1996).
\newblock D-{NURBS:} a physics-based framework for geometric design.
\newblock {\em {IEEE} Trans. Vis. Comput. Graph.}, 2(1):85--96.

\bibitem[Rogers and Adams 1976]{rogers_mathematical_1976}
Rogers, D.~F. and Adams, L.~A. (1976).
\newblock {\em Mathematical Elements for Computer Graphics}.
\newblock {MacGraw-Hill}, Inc.

\bibitem[Terzopoulos and Fleischer 1988]{terzopoulos_deformable_1988}
Terzopoulos, D. and Fleischer, K.~W. (1988).
\newblock Deformable models.
\newblock {\em The Visual Computer}, 4(6):306--331.

\bibitem[Terzopoulos and Qin 1994]{terzopoulos_dynamic_1994}
Terzopoulos, D. and Qin, H. (1994).
\newblock Dynamic {NURBS} with geometric constraints for interactive sculpting.
\newblock {\em {ACM} Trans. Graph.}, 13(2):103--136.

\end{thebibliography}

\appendix

\section{Appendix \label{appendix:A}}

Let us consider the expression:

\begin{equation}
Y\left(\mathbf{p},\dot{\mathbf{p}}\right)=\dot{M}\overset{\cdot}{\mathbf{p}}-\frac{1}{2}\left(\overset{\cdot}{\mathbf{p}}\right)^{T}\frac{\partial M}{\partial\mathbf{p}}\overset{\cdot}{\mathbf{p}},\label{simplificando-1}
\end{equation}

By the product rule we have:

\begin{equation}
\overset{\cdot}{M}=\frac{d}{dt}\left[\int_{u}\mu J^{T}Jdu\right]=\int_{u}\mu J^{T}\overset{\cdot}{J}du+\int_{u}\mu\left(\overset{\cdot}{J}\right)^{T}Jdu.\label{simplificando-2}
\end{equation}

Now, let us define the expressions $I$ and $\hat{I}$ as:

\begin{equation}
I\equiv\int_{u}\mu J^{T}\overset{\cdot}{J}du,\label{eq:compute-I}
\end{equation}

\begin{equation}
\hat{I}\equiv\int_{u}\mu\left(\overset{\cdot}{J}\right)^{T}Jdu.\label{eq:compute-Ichapeu}
\end{equation}

Therefore, we can rewrite expression (\ref{simplificando-1}):

\begin{equation}
Y\left(\mathbf{p},\dot{\mathbf{p}}\right)=I\overset{\cdot}{\mathbf{p}}+\hat{I}\overset{\cdot}{\mathbf{p}}-\frac{1}{2}\left(\overset{\cdot}{\mathbf{p}}\right)^{T}\frac{\partial M}{\partial\mathbf{p}}\overset{\cdot}{\mathbf{p}},\label{simplificando-1-1}
\end{equation}

Now, we prove that:

\begin{equation}
\hat{I}\dot{\mathbf{p}}=\left[\int_{u}\mu\left(\overset{\cdot}{J}\right)^{T}Jdu\right]\overset{\cdot}{\mathbf{p}}=\frac{1}{2}\left(\overset{\cdot}{\mathbf{p}}\right)^{T}\frac{\partial M}{\partial\mathbf{p}}\overset{\cdot}{\mathbf{p}}.\label{simplificando-3}
\end{equation}

If we name: 
\begin{equation}
R=\frac{1}{2}\left(\overset{\cdot}{\mathbf{p}}\right)^{T}\frac{\partial}{\partial p_{i}}\left(J^{T}J\right)\overset{\cdot}{\mathbf{p},}\label{eq:definition-R}
\end{equation}
then, by applying the product rule we get: 
\begin{equation}
R=\frac{1}{2}\left(\overset{\cdot}{\mathbf{p}}\right)^{T}\left(\frac{\partial J}{\partial p_{i}}\right)^{T}J\overset{\cdot}{\mathbf{p}}+\frac{1}{2}\left(\overset{\cdot}{\mathbf{p}}\right)^{T}J^{T}\frac{\partial J}{\partial p_{i}}\overset{\cdot}{\mathbf{p}}.\label{eq:EvolutionSimple04}
\end{equation}

Let $\mathbf{j}_{i}$ be the collum $i$ of the Jacobian $J$. Once
$J=J\left(\mathbf{p}\right),$ where $\mathbf{p}=\mathbf{p}\left(t\right),$
then the Chain-Rule allows to write:

\begin{equation}
\dot{\mathbf{j}_{i}}=\frac{d}{dt}(\mathbf{j}_{i})=\frac{\partial J}{\partial p_{i}}\frac{d}{dt}\left(\mathbf{p}\right)=\frac{\partial J}{\partial p_{i}}\overset{\cdot}{\mathbf{p}}.\label{eq:equation-for-R}
\end{equation}

Expression (\ref{eq:EvolutionSimple04}) can be rewritten as: 
\begin{equation}
R=\frac{1}{2}\left(\frac{\partial J}{\partial p_{i}}\overset{\cdot}{\mathbf{p}}\right)^{T}J\overset{\cdot}{\mathbf{p}}+\frac{1}{2}\left(J\overset{\cdot}{\mathbf{p}}\right)^{T}\frac{\partial J}{\partial p_{i}}\overset{\cdot}{\mathbf{p},}\label{eq:equation-for-R-modif}
\end{equation}

So, by substituting equation (\ref{eq:equation-for-R}) in expression
in (\ref{eq:equation-for-R-modif}) we obtain:

\begin{equation}
R=\frac{1}{2}\left(\dot{\mathbf{j}_{i}}\right)^{T}J\overset{\cdot}{\mathbf{p}}+\frac{1}{2}\left(J\overset{\cdot}{\mathbf{p}}\right)^{T}\left(\dot{\mathbf{j}_{i}}\right),
\end{equation}
\begin{equation}
R=\frac{1}{2}\left(\dot{\mathbf{j}_{i}}^{T}J\overset{\cdot}{\mathbf{p}}\right)+\frac{1}{2}\left(\dot{\mathbf{j}_{i}}^{T}J\overset{\cdot}{\mathbf{p}}\right)^{T}=\left(\dot{\mathbf{j}_{i}}\right)^{T}J\dot{\mathbf{p}}.\label{eq:redefinition-R}
\end{equation}

Therefore, from the expressions (\ref{eq:definition-R}) and (\ref{eq:redefinition-R})
we get that:

\begin{equation}
\left(\dot{\mathbf{j}_{i}}\right)^{T}J\dot{\mathbf{p}}=\frac{1}{2}\left(\overset{\cdot}{\mathbf{p}}\right)^{T}\frac{\partial}{\partial p_{i}}\left(J^{T}J\right)\overset{\cdot}{\mathbf{p}},\,\, for\,\, i=0,\ldots,4\left(n+1\right),\label{eq:EvolutionSimple03-1}
\end{equation}
which is equivalent to expression (\ref{simplificando-3}).

Therefore, by substitution this result in equation (\ref{simplificando-1-1})
we obtain:

\begin{equation}
Y\left(p,\dot{p}\right)=I\overset{\cdot}{\mathbf{p}},\label{simplificando-1-1-1}
\end{equation}
where $I$ is computed by expression (\ref{eq:compute-I}).

\section{Appendix \label{appendix:B}}

In order to prove that:

\begin{equation}
\left[\frac{1}{2}\mathbf{p}^{T}\frac{\partial K}{\partial\mathbf{p}}\mathbf{p}\right]=0,\label{appendixB00-1}
\end{equation}
we must observe that:

\begin{equation}
\frac{\partial\mathbf{c}}{\partial p_{i}}=\frac{\partial J}{\partial p_{i}}\mathbf{p}+\mathbf{j}_{i}.\label{appendixB00-2}
\end{equation}

However, due to the definition of jacobian matrix $J$ we must have
$\frac{\partial\mathbf{c}}{\partial p_{i}}=\mathbf{j}_{i}.$ Therefore:

\begin{equation}
\frac{\partial J}{\partial p_{i}}\mathbf{p}=0.\label{appendixB00-2-1}
\end{equation}

On the other hand:

\begin{equation}
\left[\frac{1}{2}\mathbf{p}^{T}\frac{\partial K}{\partial\mathbf{p}}\mathbf{p}\right]=\left[\frac{1}{2}\mathbf{p}^{T}\frac{\partial K}{\partial p_{0}}\mathbf{p}\,\,\frac{1}{2}\mathbf{p}^{T}\frac{\partial K}{\partial p_{1}}\mathbf{p}\ldots\frac{1}{2}\mathbf{p}^{T}\frac{\partial K}{\partial p_{i}}\mathbf{p}\ldots\frac{1}{2}\mathbf{p}^{T}\frac{\partial K}{\partial p_{4\left(n+1\right)}}\mathbf{p}\right].\label{appendixB00-1-1}
\end{equation}

But, from the definition of $K$ matrix in expression (\ref{Eint04}):

\begin{equation}
\mathbf{p}^{T}\frac{\partial K}{\partial p_{i}}\mathbf{p}=\mathbf{p}^{T}\left(\frac{\partial}{\partial p_{i}}\int_{u}\left(\alpha J_{u}^{T}J_{u}+\beta J_{uu}^{T}J_{uu}\right)du\right)\mathbf{p}.
\end{equation}

Once the vector $p$ does not depend on the parameter $u$ we can
write the first term inside the integral as:

\begin{equation}
\alpha\mathbf{p}^{T}\left(\int\left(\frac{\partial J_{u}^{T}}{\partial p_{i}}J_{u}+J_{u}^{T}\frac{\partial J_{u}}{\partial p_{i}}\right)du\right)\mathbf{p}=
\end{equation}

\begin{equation}
\alpha\mathbf{p}^{T}\left(\int\frac{\partial J_{u}^{T}}{\partial p_{i}}J_{u}du\right)\mathbf{p}+\alpha\mathbf{p}^{T}\left(\int J_{u}^{T}\frac{\partial J_{u}}{\partial p_{i}}du\right)\mathbf{p}=
\end{equation}

\begin{equation}
\alpha\mathbf{p}^{T}\left(\int\frac{\partial}{\partial u}\left(\frac{\partial J^{T}}{\partial p_{i}}\right)J_{u}du\right)\mathbf{p}+\alpha\mathbf{p}^{T}\left(\int J_{u}^{T}\frac{\partial}{\partial u}\left(\frac{\partial J}{\partial p_{i}}\right)du\right)\mathbf{p}=0,
\end{equation}
\begin{equation}
\alpha\left(\int\frac{\partial}{\partial u}\left(\frac{\partial J}{\partial p_{i}}\mathbf{p}\right)^{T}J_{u}du\right)\mathbf{p}+\alpha\left(\int J_{u}^{T}\frac{\partial}{\partial u}\left(\frac{\partial J}{\partial p_{i}}\mathbf{p}\right)du\right)=0,
\end{equation}
due to equation (\ref{appendixB00-2-1}). Therefore, expression (\ref{appendixB00-1})
has been proved.
\end{document}